\documentclass[aps,prd,a4paper,10pt,superscriptaddress,preprintnumbers]{revtex4}

\usepackage{enumerate}
\usepackage{amsmath,amssymb,bm}
\usepackage[dvipdfmx]{graphicx}
\usepackage{braket}
\usepackage{url} 
\usepackage{color} 

\usepackage{layout}
\allowdisplaybreaks[4]

\def\msol{M_{\odot}}
\def\mbh{M_{\rm BH}}
\def\meff{M_{\rm eff}}
\def\rhosp{\rho_{\rm sp}}
\def\rsp{r_{\rm sp}}

\def\rmin{r_{\rm min}} 
\def\a{\alpha}

\def\mdm{M_{\rm DM}}

\def\bh{{\rm BH}}

\def\vir{{\rm vir}}
\def\cri{{\rm cri}}

\def\sourcedistance{D}

\def\pc{{\rm pc}}
\def\km{{\rm km}}

\def\apj{Astrophys. J.}
\def\apjl{Astrophys. J. Lett.}
\def\cqg{Classical Quant. Grav.}
\def\pasj{Publ. Astron. Soc. Jpn}
\def\mnras{Mon. Not. R. Astron. Soc.}

\begin{document}

\title{
Gravitational waves as a probe of dark matter mini-spikes
%
%
%
%
%
%
%
}
\author{Kazunari Eda}{1}
\email{eda@resceu.s.u-tokyo.ac.jp}
\affiliation{
Department of Physics, 
Graduate School of Science, 
The University of Tokyo, Tokyo, 113-0033, Japan
}
\affiliation{
Research center for the early universe (RESCEU), 
Graduate School of Science, 
The University of Tokyo, Tokyo, 113-0033, Japan
}
\author{Yousuke Itoh}
\affiliation{
Research center for the early universe (RESCEU), 
Graduate School of Science, 
The University of Tokyo, Tokyo, 113-0033, Japan
}
\author{Sachiko Kuroyanagi}
\affiliation{
Department of Physics, Faculty of Science, Tokyo University of Science, 1-3, 
Kagurazaka, Shinjuku-ku, Tokyo 162-8601, Japan
}
\author{Joseph Silk}
\affiliation{
Institut d' Astrophysique, UMR 7095, CNRS, UPMC Univ. Paris VI, 98 bis Boulevard Arago, Paris 75014, France
}
\affiliation{
Department of Physics and Astronomy, The Johns Hopkins University, Baltimore, MD 21218, USA}
\affiliation{
Beecroft Institute for Particle Astrophysics and Cosmology, Department of Physics, University of Oxford, Keble Road, Oxford, OX1 3RH, UK
} 

\begin{abstract}
  Recent studies show that an intermediate mass black hole (IMBH) may
  develop  a dark matter (DM) mini-halo according to some BH
  formation scenarios.
  We consider a binary system composed of an IMBH
  surrounded by a DM mini-spike and a stellar mass object orbiting
  around the IMBH. The binary evolves due to gravitational pull and
  dynamical friction from the DM mini-spike and back-reaction from its
  gravitational wave (GW) radiation which can be detected by future
  space-borne GW experiments such as eLISA/NGO.  We consider a single
  power-law model for the DM mini-spike which is assumed to consist of
  non-annihilating DM particles and demonstrate that an eLISA/NGO
  detection of GW from such a binary enables us to measure the DM
  mini-spike parameters very accurately.  For instance, in our
  reference case originally advocated by
  \citeauthor{Zhao:2005zr} (\citeyear{Zhao:2005zr}) and
  \citeauthor{Bertone:2005xz} (\citeyear{Bertone:2005xz}), we could
  determine the power-law index $\alpha$ of the DM mini-spike radial
  profile with a 1 $\sigma$ relative error of $\pm 5\times 10^{-6}$
  for a GW signal with signal-to-noise-ratio 10 and assuming a 5 year
  observation with eLISA. We also investigate how accurately the DM
  parameters can be determined for various DM parameters and the
  masses of the IMBH-stellar mass object binary surrounded by a DM
  mini-spike.  We find that we can determine the power-law index
  $\alpha$ at 10 \% level even for a slightly flatter radial
  distribution of $\alpha \sim 1.7$.
\end{abstract} 
  
\preprint{RESCEU-41/14}

\maketitle  

\section{Introduction} 
There is much reliable evidence for the existence of dark matter (DM)
which is mainly associated with the missing mass problem.  Astronomers and
particle physicists seek to probe DM properties
by direct laboratory experiments or indirect observations
\cite{Arrenberg2013}.  Indirect techniques include efforts
to detect gamma rays from DM annihilation using telescopes
such as the Fermi Large Area Telescope (Fermi-LAT, \cite{Fermi2013}),
the Major Atmospheric Gamma-ray Imaging Cherenkov (MAGIC) telescope
\cite{MAGIC2008}, the High Energy Stereoscopic System (H.E.S.S.,
\cite{HESS2013}) and the Very Energetic Radiation Imaging Telescope
Array System (VERITAS, \cite{VERITAS2012}) (see, e.g., \cite{Funk2013}
for a review).

It was first suggested by Gondolo and Silk \cite{Gondolo:1999ef} that
adiabatic growth of a BH at the center of a DM halo whose density had
a singular power-law cusp $\rho\left(r\right) \propto
r^{-\alpha_{\text{ini}}}$ with $0\leq \alpha_{\text{ini}} \leq 2$ led
to a high density DM region around the central BH, $\rho_{\rm
  spike}\left(r\right) \propto r^{-\alpha}$ with $2.25\leq \alpha \leq
2.5$.  This region is called a DM spike.  Inside the spike, DM
annihilations are enhanced and produce the strong gamma-ray photon flux
which could be detectable to the telescopes mentioned above.

However, subsequent studies pointed out that this spike could be
weakened by dynamical processes such as mergers of host-galaxies,
sub-halo accretion and passing of molecular clouds
\cite{NakanoMakino1999,Ullio:2001fb,Merritt:2002vj,Merritt:2003qk,Bertone:2005hw}.
These processes transfer energy to the DM particles and destroy the
structure of the DM spike.  Then the annihilation rate in the spike is
smaller than predicted in \cite{Gondolo:1999ef} because it depends on
the line-of-sight integral of the squared density of the spike.  If
supermassive black holes (SMBHs)  have experienced mergers,
they are unlikely to have surviving spike structure.  Even this however is controversial  because of the uncertainty in whether the final parsec problem for SMBH mergers has been resolved phenomenologically \cite{Liu:2013fca}
or even theoretically \cite{Vasiliev:2013nha}.
On
the other hand, formation scenarios of intermediate-mass black holes
(IMBH) which allow DM mini-spikes have been proposed
\cite{Zhao:2005zr,Bertone:2005xz}.  If the IMBH have never experienced
mergers in the past, the DM mini-spike around the central IMBH
is likely to survive.

IMBHs may exist in our universe \cite{2012Sci...337..554W}, and even
several hundreds would reside in the halo of the present-day Milky Way
galaxy \cite{2003MNRAS.340..647I,2014ApJ...780..187R}. Those IMBHs in
globular clusters are recognized as promising sources for the evolved
Laser Interferometer Space Antenna (eLISA) \cite{2007CQGra..24R.113A}
/ the New Gravitational Wave Observatory (NGO)
\cite{AmaroSeoane:2012km} and DECi-hertz Interferometer Gravitational
Wave Observatory (DECIGO)\cite{Kawamura:2011zz}.

In our previous work \cite{Eda:2013gg}, we demonstrated that a very
tiny effect such as the  gravitational pull of a DM mini-spike around an IMBH
indeed affects detectability of GW by eLISA and thereby we could infer
presence or absence of a DM mini-spike around an IMBH using GW.
Specifically, when a stellar mass object inspirals into the central 
IMBH, it is affected by the gravitational force of not only the
central IMBH but also the mini-spike.  Therefore the inspiral GW is
modified by the mini-spike around the central IMBH.  We found
that the very tiny effect from gravitational pull of a DM mini-spike
could have a large impact on detectability of the GW, thanks to the
huge number of orbital cycles which the binary experienced in the
eLISA detection frequency band.  We also found that GW detectability
strongly depends on the density profile of the DM mini-spike.

In this paper, we extend our previous work in the following way.  We
again consider GWs emitted from a binary system consisting of a
stellar mass object and an IMBH harbored in a DM mini-spike, and
calculate the GW waveform including the effect of both the
gravitational potential and the dynamical friction on the falling
stellar mass object in the DM mini-spike.  Furthermore, we investigate
how accurately the DM parameters are determined by the GW
observations.  We find that the DM information contained in the
waveform can be extracted with very good accuracy by GW observations
if the central IMBH has a steep density mini-spike.  We also
investigate how the detection accuracy of the DM parameters changes
depending on the masses of the binary components and the density
profile of the DM mini-spike such as the power index and
overall normalization.

Recently \citeauthor{Macedo:2013qea} made clear the importance of the
dynamical friction on the GW waveform in a quite different context from
ours, namely, a stellar mass object falling in a compact configuration
of DM clouds \cite{Macedo:2013qea}. Also,
\citeauthor{2014arXiv1404.7140B} has given a wide survey on
astrophysical environmental effects on GW signals using order of
magnitude estimates, concluding that astrophysical environmental
effects such as accretion disks, magnetic fields, and DM halos do not
obscure gravitational wave astrophysics, e.g., precision measurements
of binary masses and tests of general relativity
\cite{2014arXiv1404.7140B} (See also \cite{2014PhRvD..89j4059B}).  To indicate
one exception, our paper shall clearly show that, in the recently
advocated DM mini-spike scenario, environmental effects do affect GW
detectability \cite{Eda:2013gg} and we can measure DM properties quite
accurately from eLISA GW detection, which will be shown through a
detailed study using a matched filtering technique and Fisher matrix
analysis.

We stress that while gamma-ray observations from DM annihilations can
only work if the DM is a weakly interacting massive particle (WIMP), the GW
observations we proposed should be widely applicable to any type of DM
particles.  Furthermore, matter is almost completely transparent to
GWs unlike electromagnetic waves because of the smallness of their
gravitational cross-section.  Hence, the GWs carry pure information on
the DM from the mini-spike to the detector.  Future GW experiments
will probe the structure of the DM mini-spike and will even offer a
hint on the nature of the DM particle.

The rest of the paper is organized as follows.
Sec. \ref{Sec:mini-halo_model} presents the DM mini-spike model and
candidates for the stellar mass object.  In Sec. \ref{Sec:GW_waveform}
we derive the GW waveform from the system which we consider and the
observational errors of the waveform parameters are calculated in
Sec. \ref{Sec:parameter_resolution}.  Finally our conclusions are given in
Sec. \ref{Sec:conclusion}.

\section{Mini-halo model}\label{Sec:mini-halo_model}
\subsection{Initial DM mini-halo profile}
We assume that the initial DM mini-halo profile which 
leads to the DM mini-spike after the adiabatic growth of the IMBH 
is approximately described by the Navarro-Frenk-White (NFW) profile \cite{1997ApJ...490..493N}
\begin{align}
 \rho_{\text{NFW}}\left(r\right)=\dfrac{\rho_s}{\left(r/r_s\right)\left(1+r/r_s\right)^2}, 
\end{align}
where $r$ is the radius, $\rho$ is the mass density and the subscript
``s'' stands for the scaling.  Navarro, Frenk and White obtained this
profile via cosmological N-body numerical simulation and numerical
fitting of obtained DM halo profiles around clusters of
galaxies. Surprisingly, their simulations showed that every DM halo
around a cluster of galaxies follows the NFW profile when normalized
properly by $\rho_s$ and $r_s$. Later work, however, shows that the
inner slope may be slightly steeper than the NFW one ($\rho\propto
r^{-1}$) and may not even be universal \cite{FukushigeKawaiMakino2004}.  In
any case, we refer to the NFW profile and the derived parameters
listed in the table \ref{tbl:DMParameters} below as our reference
model in the following for simplicity. We study how accurately the
dark matter parameters can be measured in our reference model in
detail, then extend our analysis to different sets of values of
the DM parameters to take into account ambiguities in the DM distribution
around an IMBH.

The NFW parameters $\rho_s$ and $r_s$ are related to the cluster mass
and concentration parameters by 
\begin{subequations}
\begin{align}
 &M_{\vir} =
 \dfrac{4\pi}{3}\Delta_{\vir}\Omega_m \left(z_f\right)\rho_{\cri}\left(z_f\right)r_{\vir}^3,
\label{eq:Mvir}
\\
 &\rho_s \equiv \dfrac{1}{3 f\left(c_{\vir}\right)}\Delta_{\vir}\Omega_m \left(z_f\right)\rho_{\cri}\left(z_f\right)
    c_{\vir}^3, 
\label{eq:rhos}
\end{align}
\end{subequations}
where $c_{\vir} \equiv r_{\vir}/r_s$ and $r_{\vir}$ is the virial
radius and $M_{\vir}$ is the virial mass of the cluster, $z_f$ is the
formation redshift of the cluster, $\Omega_m$ is the matter density
parameter, $\rho_{\cri}$ is the critical matter density of the
universe and the function $f\left(x\right)$ is the volume integral of
the NFW profile $f\left(x\right) \equiv
\ln\left(1+x\right)-x/\left(1+x\right)$ (see, e.g.,
\cite{Okabe:2007af}).  We used the fitting formula given by
\cite{2001ApJ...559..572O} for the parameter $\Delta_{\vir}$:
$\Delta_{\vir} \equiv 18\pi^2(1+0.4093 \omega_{\vir}^{0.9052})$ where
$\omega_{\vir}\equiv 1/\Omega_{m}(z_{f})-1$
\cite{2001ApJ...559..572O}.  The mass-concentration relation is taken
from \cite{Duffy:2008pz} which fits the profiles of the clusters of
galaxies obtained in their N-body simulations.  This  result for
clusters of galaxies may or may not apply for the
mini-halo. In any case, concentration parameters between $O(1-1000)$
lead to qualitatively similar results and are given by the following
relation.
\begin{align}
c_{200} &= A_{200} \left( M_{200}/M_{\rm pivot} \right)^{B_{200}}(1+z_f)^{C_{200}},
\label{eq:concentration}
\end{align} 
where we assume $(A_{200},B_{200},C_{200},M_{\rm pivot}) =
(5.71,-0.084,-0.47, 1.0\times 10^{14}h^{-1}M_{\odot})$ from the result
of \cite{Duffy:2008pz}. The parameters $A_{200}$ and so on may be
appropriately used when the overdensity $\Delta_{\vir}$ equals $200$.
However, we here assumed $A_{200}\simeq A_{\vir}$ and so on for
simplicity.

As will be shown later, the GW waveform depends on the DM mini-spike slope
$\alpha$ and some combination of the radius at which the mini-spike
is established, $\rsp$, and the DM density there, $\rhosp$.  Under the
assumption of adiabatic growth, while the final power-law index of the
DM mini-spike, $\alpha$, depends on the power-law index of the initial
inner DM profile, the latter two depend on $\rho_s$, $r_s$ and
$\alpha$.  For concreteness, we adopt $M_{\vir} = \mdm = 10^6\msol$,
$z_f=20$ in Eqs. (\ref{eq:Mvir}), (\ref{eq:rhos}), and
(\ref{eq:concentration}) \cite{Zhao:2005zr,Bertone:2005xz} and find
$c_{\vir} = 6.6$, $r_s = 23.1\pc$, and $\rho_s = 3.8\times10^{-22}
{\rm g/cm}^3$.

\subsection{DM mini-spike profile}
We proceed to discuss the DM profile of the mini-spike.  If the DM
mini-halo initially has a cuspy profile $\rho\left(r\right) \propto
r^{-\alpha_{\text{ini}}}$ with $0 \leq \alpha_{\text{ini}} \leq 2$,
then the adiabatic growth of the central IMBH produces the DM
mini-spike.  Hence the dark matter profile becomes
\cite{Gondolo:1999ef,Quinlan:1994ed}
\begin{align}
  \rho_{\text{DM}} \left(r \right) = 
    \begin{cases}
      \rho_{\rm spike}\left(r\right),  & \left( r_{\min} \leq r \leq \rsp \right), \\
      \rho_{\text{NFW}}\left(r\right), & \left(  \rsp < r\right), 
    \end{cases}
\end{align}
with
\begin{subequations}
\begin{align}
 &\rho_{\rm spike}\left(r\right) = \rhosp \left(\dfrac{\rsp}{r}\right)^{\alpha}, 
\label{eq:MiniSpikeDensity}\\
 &\alpha = \dfrac{9-2\alpha_{\text{ini}}}{4-\alpha_{\text{ini}}},
\end{align}
\end{subequations} 
where $\rhosp$ is the normalization constant and $\rsp$ is empirically
defined by $\rsp \sim 0.2 r_h$.  The radius $r_h$ is the distance of
the gravitational influence of the central IMBH with the mass
$M_{\bh}$ and is approximately obtained by $M\left(<r_h\right) =
4\pi\int_{0}^{r_h}\rho_{\text{DM}} \left(r\right) r^2 dr = 2M_{\bh}$
\footnote{ In \cite{Eda:2013gg}, we have used $\rho_{\text{NFW}}$
  instead of $\rho_{\text{DM}}$ to estimate $\rhosp$ and $\rsp$.}.
The slope of the DM mini-spike takes the value $2.25 \leq \alpha \leq
2.5$ for $0 \leq \alpha_{\text{ini}} \leq 2$.  In the case of an
initially NFW profile, $\alpha_{\text{ini}} = 1$, this gives rise to
$\alpha = 7/3$.  If the initial profile of the mini-halo is a uniform
distribution, then the final profile after the adiabatic growth of the
IMBH would become a more gentle $\rho_{\rm spike}\left(r\right)
\propto \left(r/r_h\right)^{-3/2}$
\cite{1980ApJ...242.1232Y,1995ApJ...440..554Q,Ullio:2001fb}.

It is important to note that the final profile of the DM mini-halo
depends on the formation history of the central IMBH.  If the IMBH has
experienced disruptive processes such as mergers in the past, the
mini-spike would be weakened or disappear.  For this reason, we do not
specify the value of the power-law index $\alpha$ of the DM mini-spike
and treat it as a free parameter within the range $0\le \alpha\le 3$.
In the following, even if $\alpha < 2.25$, we will still call the DM
distribution close to the central IMBH described by
Eq. (\ref{eq:MiniSpikeDensity}) ``a DM mini-spike'' for the sake of simplicity.
Indeed we will see that the ``DM mini-spike'' leaves its
signature in the GW waveform when $\alpha \gtrsim 1.7$, but certainly does not
when $\alpha = 0$.  We will also assume different values of $\rhosp$
to study how the ambiguities in $\rho_s$ and $r_s$ mentioned above
affect our results.  Finally, we take $r_{\min}$ to be the innermost
stable circular orbit of the central IMBH, $r_{\min} = r_{\rm ISCO}
\equiv 6G\mbh/c^2$. It may be more precise to use $4G\mbh/c^2$
\cite{Sadeghian:2013laa}, but such a change of $r_{\min}$ does not
alter at all the measurement accuracy of the DM parameters shown
below.  The parameters of the DM density profile are summarized in the
table \ref{tbl:DMParameters} below.

\begin{table}[hpt]
\begin{tabular}{|c|c|c|c|c|}
\hline
 $\mdm$ & $\mbh$ & $z_f$ & $c_{\rm halo}$ & $r_{\vir}$  \\
\hline
 $10^6\msol$ & $10^3\msol$ & $20$ & $6.6$ & $152.6\pc$  \\
\hline
\hline
 $r_s$ & $\rho_s$ & $r_h$ & $\rsp$ & $\rhosp$    \\
\hline
 $23.1\pc$ & $3.8\times10^{-22}{\rm g/cm}^3$ & $1.65\pc$ & $0.54\pc$ &$226\msol/pc^3$    \\
\hline
\end{tabular}
\caption{
Our reference 
model parameters of the IMBH, 
the DM mini-halo and the DM mini-spike.
$\mdm$: The total mass of the mini-halo, 
$\mbh$: the mass of the central intermediate mass black hole,
$z_{f}$: the formation redshift of the mini-halo,
$c_{\rm halo}$: the concentration of the mini-halo,
$r_{\vir}$: the virial radius of the mini-halo,
$r_s$: the NFW $r_s$ parameter of the mini-halo,
$\rho_s$: the NFW $\rho_s$ parameter of the mini-halo,
$r_h$: the radius at which $\mdm(r_h)=2\mbh$,
$\rsp$: the radius where the spike forms (estimated by $\rsp = 0.2r_h$), and 
$\rhosp$: the mini-halo mass density at $\rsp$.}
\label{tbl:DMParameters}
\end{table}

\subsection{Candidate for a stellar mass object}

Before moving onto the calculation of the GW waveform, we discuss what can
be a candidate for ``a stellar mass object''.  Let us consider a
stellar mass object with mass $\mu$ denoted by $A$, orbiting around an
intermediate mass black hole $B$ with mass $\mbh$.  We consider the
inspiral up to the the innermost stable circular orbit $r_{\rm ISCO}$
\begin{align}
 r_{\rm ISCO}&= \dfrac{6G\mbh}{c^2} \simeq 9\times 10^3 \km
\left(\dfrac{\mbh}{10^3\msol}\right).
\end{align}
Hence, the object $A$ should have a radius smaller  than at most $9\times
10^3$ km.  At the same time, the tidal radius of $A$ orbiting $B$ at
the orbital radius of $r_{\rm ISCO}$ is
\begin{align}
 l_{\rm A~tidal}
&\simeq r_{\rm ISCO}\left(\dfrac{\mu}{\mbh}\right)^{1/3} 
\simeq 
9\times 10^2 \km
\left(\dfrac{\mu}{1\msol}\right)^{1/3}\left(\dfrac{\mbh}{10^3\msol}\right)^{-1/3}
\left(\dfrac{\mbh}{10^3\msol}\right)
\end{align}
Hence, this object must be either a black hole or a neutron star.
Alternatively, if we assume $A$ to be a white dwarf of radius $l_A =
10000$km or a sun-like object of radius $l_A = 10^6\km$, the innermost
orbital radius should be replaced by the radius below which the object
$A$ is tidally destroyed
\begin{align} 
 r_{\rm tidal}&\simeq
\left(\dfrac{\mbh}{\mu}\right)^{1/3} l_{A} 
\simeq 3\times 10^{-7}\pc
\left(\dfrac{\mbh}{10^3\msol}\right)^{1/3}
\left(\dfrac{\mu}{1\msol}\right)^{-1/3}
\left(\dfrac{l_{A}}{10^6\km}\right).  
\end{align}
As will be stated, we will consider the orbital radius of order
$10^{-8}\pc$ or less, so we cannot assume our stellar mass object to
be a normal star with radius $\sim 10^6$\km. A white dwarf may be
an interesting candidate since an electromagnetic counterpart may be
expected when it is tidally disrupted (e.g.,
\cite{2012ApJ...749..117H,2013ApJ...769...85S,2014arXiv1405.1426M}). Yet,
here in this paper we assume a neutron star or a black hole when we
refer to a stellar mass object.

\section{GW waveform}\label{Sec:GW_waveform}
\subsection{Equation of motion for the stellar mass object}
Let us consider a binary system which involves a small compact object
with a mass of $\mu = 1 M_{\odot}$ and an IMBH with a mass of
$M_{\text{BH}} = 10^3 M_{\odot}$.  The mass of the stellar mass object
$\mu$ is much smaller than the mass of IBMH $M_{\text{BH}}$.  So the
reduced mass is approximately equal to $\mu$ and the barycenter
position is approximately equal to the position of the IMBH.  By
adopting a reference frame attached to the barycenter, the equation of
motion of the radial relative separation between the stellar mass
object and the IMBH describes the motion of the former and is given by
\begin{align}
 \dfrac{d^2 r}{dt^2} &= - \dfrac{G \meff}{r^2} - \dfrac{F}{r^{\alpha -1}} + \dfrac{h^2}{r^3}, \label{Eq:EOM} 
\end{align}
where $h$ is the angular momentum of the stellar mass object per its
mass, and $\meff$ and $F$ are defined by
\begin{subequations}
\begin{align}
  &\meff =
    \begin{cases}
      \mbh - M_{\text{DM}} \left( < r_{\text{min}}\right)
& \left( \rmin \leq r \leq \rsp \right), \\
      \mbh & \left( r < \rmin \right),
    \end{cases} \\
  &F =
    \begin{cases}
      \rmin^{\alpha-3} M_{\text{DM}} \left( < r_{\text{min}}\right)  &  \left( \rmin \leq r \leq \rsp \right), \\
      0 &  \left( r < \rmin \right).
    \end{cases}
\end{align}
\end{subequations}
The mass $M_{\text{DM}} \left( < r_{\text{min}}\right)$ denotes the DM
mass contained within the ISCO and is defined as $M_{\text{DM}} \left(
  < r_{\text{min}}\right) \equiv 4\pi \rsp^\alpha \rhosp
\rmin^{3-\alpha}/\left(3-\alpha\right)$.  The first term on the
right-hand side of Eq. (\ref{Eq:EOM}) describes the gravitational
potential of the effective mass of the central IMBH which is modified
by the DM due to the absence of the DM within the ISCO, the second
term accounts for the DM effect, and the third term represents a
centrifugal force.  Here the dynamical friction force and the GW back
reaction force are neglected because these effects are much smaller
than the gravitational potential of the IMBH. We will introduce these
effects to include an adiabatic evolution of the orbital radius in the
next subsection.

We assume that the stellar mass object orbits in a circular manner for
simplicity.  The orbital radius $R$ is obtained by solving
$d^2r/dt^2=0$ in Eq. (\ref{Eq:EOM}).  The orbital frequency $\omega_s$
is related to the angular momentum $h$ by $R\omega_s$,
so we get
\begin{align}
  \omega_s &= \left[ \dfrac{GM_{\text{eff}}}{R^3} + \dfrac{F}{R^{\alpha}} \right]^{1/2}. \label{Eq:orbital_frequency}
\end{align}
When a DM mini-spike is not present around the IMBH, $F \to 0$ and
$\meff \to \mbh$ , so Eq. (\ref{Eq:orbital_frequency}) leads to the
Kepler's law $\omega_s^2 = G\mbh/R^3$.

\subsection{Energy balance equation}
In this subsection, we introduce the GW back-reaction and the
dynamical friction into the stellar mass object's orbit by taking the
energy balance equation into account.  When the stellar mass object
orbits around the IMBH, a part of its energy $E_{\text{orbit}}$ is
converted into  GW emission loss $E_{\text{GW}}$ and dynamical
friction loss $E_{\text{DF}}$.  Thus the following energy balance
equation is satisfied:
\begin{align}
 -\dfrac{d E_{\text{orbit}}}{dt} = \dfrac{d E_{\text{GW}}}{dt} + \dfrac{d E_{\text{DF}}}{dt}. \label{Eq:energy_balance_eq}
\end{align}
As we will see in this subsection, this energy balance equation gives
the time evolution of the orbital radius.  The resulting orbit can be
regarded as a quasi-circular orbit because of the smallness of these
dissipative effects.

The orbital energy $E_{\text{orbit}}$ is the sum of the kinetic energy
and the gravitational potential of the stellar mass object, so we can
calculate $E_{\text{orbit}}$ using Eq. (\ref{Eq:orbital_frequency}),
\begin{align}
E_{\text{orbit}}
  &= \dfrac{1}{2} \mu v^2 + \dfrac{h^2}{2R^2} - \dfrac{G\mu M_{\text{eff}}}{R} \nonumber \\
  &= -\dfrac{G\mu M_{\text{eff}}}{2R} + \dfrac{4-\alpha}{2\left( 2- \alpha\right)} \dfrac{\mu F}{R^{\alpha-2}}, \label{Eq:orbital_energy} 
\end{align}
where $v$ is the orbital velocity.  When we consider the evolution of
the radius $R$, $dR/dt$ does not vanish. So the time derivative of
Eq. (\ref{Eq:orbital_energy}) gives the following equation,
\begin{equation}
\dfrac{d E_{\text{orbit}}}{dt} = \left( \dfrac{GM_{\text{eff}}}{2R^2}  + \dfrac{4-\alpha}{2}\dfrac{F}{R^{\alpha - 1}} \right)\mu \dfrac{dR}{dt}. \label{Eq:E_orbit}
\end{equation}
To the lowest order in the Post Newtonian expansion, the gravitational
radiation energy is given by the quadrupole formula.  We apply the
formula to the circular Newtonian binary and obtain
\begin{equation}
 \dfrac{d E_{\text{GW}}}{dt} = \dfrac{32}{5} \dfrac{G\mu^2}{c^5} R^4 \omega_s^6 .  \label{Eq:E_GW}
\end{equation}
When the stellar mass object moves through the cloud of  DM, it
gravitationally interacts with DM particles.  This effect is called
dynamical friction, sometimes referred to as  gravitational
drag which was first discussed by Chandrasekar
\cite{1943ApJ....97..255C}.  Because of dynamical friction, the
stellar mass object running through the DM halo is decelerated in the
direction of its motion and loses its kinetic energy as well as its
angular momentum.  The dynamical friction force is given by $f_{\text{DF}}
= 4\pi G^2\mu^2 \rho_{\text{DM}}(r) \ln \Lambda /v^2$ where $v$ is the
velocity of the stellar mass object \cite{2008gady.book.....B}.  The
Coulomb logarithm $\Lambda$ is defined by $\lambda \cong
b_{\text{max}} v_{\text{typ}}^2 / \left(G\mu\right)$ where
$b_{\text{max}}$ is the maximum impact parameter and $v_{\text{typ}}$
is the typical velocity of the stellar mass object.  We take $\ln
\Lambda \cong 3$. From the expression of the dynamical friction force,
we obtain the friction loss,
\begin{align}
 \dfrac{d E_{\text{DF}}}{dt} &= v f_{\text{DF}} = 4\pi G^2 \dfrac{\mu^2 \rho_{\text{DM}}(r)}{v} \ln \Lambda. \label{Eq:E_DF}
\end{align}

To find the numerical solution of the energy balance equation
(\ref{Eq:energy_balance_eq}) easily, we introduce a dimensionless
radius parameter $x$ defined by
\begin{align}
 x \equiv \varepsilon^{1/\left(3-\alpha\right)} R, \label{Eq:def_x} 
\end{align}
with 
\begin{align}
 \varepsilon \equiv \dfrac{F}{GM_{\text{eff}}}. \label{Eq:def_epsilon}
\end{align}
Using the above definition of $x$, the energy balance equation
(\ref{Eq:energy_balance_eq}) can be rewritten in the form of the
differential equation of $x$ with respect to time $t$ as
\begin{align}
  \dfrac{dx}{dt} 
   &= -c_{\text{GW}} \dfrac{\left(1+x^{3-\alpha} \right)^3}{ 4x^3\left[ 1+\left(4-\alpha \right)x^{3-\alpha} \right]}
      - c_{\text{DF}}\dfrac{1}{\left(1+x^{3-\alpha}\right)^{1/2}\left[1+\left(4-\alpha \right)x^{3-\alpha}\right]x^{- 5/2 + \alpha}}, \label{Eq:dxdt}
\end{align}
where the coefficients are defined by
\begin{subequations}
\begin{align}
  c_{\text{GW}} &\equiv \dfrac{256}{5}\left(\dfrac{G\mu}{c^3}\right)\left(\dfrac{GM_{\text{eff}}}{c}\right)^2 \varepsilon^{4/(3-\alpha)}, \label{Eq:c_GW}\\
  c_{\text{DF}} &\equiv \left(8\pi G^2\mu \rhosp \rsp^\alpha \ln \Lambda \right)\left(GM_{\text{eff}} \right)^{-3/2} \varepsilon^{\left(2\alpha-3\right)/\left[2\left(3-\alpha \right)\right]}. \label{Eq:c_DF}
\end{align}
\end{subequations}
The coefficient $c_{\text{GW}}$ is related to the gravitational
radiation energy and the coefficient $c_{\text{DF}}$ is related to the
dynamical friction.  In the case of the initially NFW profile, $\alpha
=7/3$, the coefficients $c_{\text{GW}}$ and $c_{\text{DF}} $ are $
c_{\text{GW}} = 2.0 \times 10^{-33} \ \left[1/\text{year}\right],
c_{\text{DF}} = 2.1 \times 10^{-8} \ \left[1/\text{year}\right]
$. Note that the dynamical friction coefficient $c_{\text{DF}}$ is
much larger than the gravitational radiation coefficient
$c_{\text{GW}}$.

\subsection{GW waveform}
The GW waveform from the binary composed of two compact objects with
masses $\mu$ and $\mbh$ is given by
\begin{subequations}
\begin{align}
 h_{+}\left(t\right)
  &= \dfrac{1}{\sourcedistance} \dfrac{4G\mu \omega_s^2 R^2}{c^4} \dfrac{1 + \cos ^2 \iota}{2} \cos\left(\omega_{\text{GW}}t\right), \label{Eq:GWwaveform_plus_without_backreaction} \\
 h_{\times}\left(t\right)
  &= \dfrac{1}{\sourcedistance} \dfrac{4G\mu \omega_s^2 R^2}{c^4}  \cos
 \iota \sin\left(\omega_{\text{GW}}t\right),
 \label{Eq:GWwaveform_cross_without_backreaction}
\end{align} 
\end{subequations}
where $\sourcedistance$ is the distance to the source (luminosity
distance for a cosmological source), $R$ is the orbital radius,
$\iota$ is the inclination angle, which is the angle between the
line-of-sight and the rotational axis of the orbits, and
$\omega_{\text{GW}}$ is the GW frequency which is given by
$\omega_{\text{GW}} \equiv 2 \omega_s$\cite{Maggiore:2007}.

The waveforms Eqs. (\ref{Eq:GWwaveform_plus_without_backreaction}) and
(\ref{Eq:GWwaveform_cross_without_backreaction}) are derived on the
assumption that the motion of the source is described by a circular
Newtonian orbit.  But in fact, the radius $R$ and the frequency
$\omega_s$ are not constant because the orbital energy
$E_{\text{orbit}}$ decreases gradually due to both dynamical friction
and the GW back-reaction. Including these effects, the orbit shrinks
adiabatically and becomes a quasi-circular orbit.  So the radius $R$
and the frequency $\omega_s$ should be replaced by $R \to
R\left(t\right)$, $\omega_s \to \omega_s \left(t\right)$ and the phase
$\omega_{\text{GW}}t $ should also be replaced by $\omega_{\text{GW}}t
\to \Phi\left(t\right)$ as defined by Eq. (\ref{Eq:phase_t_def})
below. Thus, the GW waveform is expressed by
\begin{subequations}
\begin{align}
 h_{+}\left(t\right)
  &= \dfrac{1}{\sourcedistance} \dfrac{4G\mu \omega_s(t)^2 R(t)^2}{c^4} \dfrac{1 + \cos ^2 \iota}{2} \cos\left[ \Phi(t)\right], \label{Eq:GWwaveform_plus_including_backreaction} \\
 h_{\times}\left(t\right)
  &= \dfrac{1}{\sourcedistance} \dfrac{4G\mu \omega_s(t)^2 R(t)^2}{c^4}  \cos \iota \sin\left[ \Phi(t)\right], \label{Eq:GWwaveform_cross_including_backreaction} \\
 \Phi \left(t\right)
  &\equiv \int^t \omega_{\text{GW}} \left(t'\right) \ dt'. \label{Eq:phase_t_def}
\end{align}
\end{subequations}

In order to discuss detectability and parameter accuracy in GW
observations, it is convenient to work in the frequency domain.
The Fourier transformation of the GW waveform is given by
\begin{align}
 \tilde{h}_{+,\times}  \left(f\right) = \int_{-\infty}^{\infty} h_{+, \times} \left(t\right) e^{2\pi ift}dt,
\end{align}
where $f$ is the GW frequency.  For simplicity, we consider a GW
coming in the detector from the optimal direction for $+$ mode.  In
such a situation, detector pattern function are $F_+ = 1 $ and
$F_\times =0 $.  So the response of the detector to the GW is $h\left(
  t\right) = h_{+}\left(t\right)$.  Using
Eq. (\ref{Eq:GWwaveform_plus_including_backreaction}), we rewrite the
GW waveform as
\begin{subequations}
\begin{align}
  &h \left(t \right) = A\left(t_{\text{ret}} \right) \cos \Phi\left(t_{\text{ret}}\right), \\
  &A\left(t\right) \equiv
    \dfrac{1}{\sourcedistance} \dfrac{4G\mu \omega_s^2 \left(t\right) R^2\left(t\right)}{c^4} \dfrac{1+\cos^2 \iota}{2}, \label{Eq:amplitude_t_def}
\end{align}
\end{subequations}
where $A\left( t\right)$ is the time-dependent amplitude and
$\Phi\left(t\right)$ is the time-dependent GW phase. In the above
equations, we have introduced the retarded time $t_{\text{ret}} \equiv
t-D/c$.  In the range of frequency we are concerned with, the
time-dependent amplitude $A\left(t\right)$ varies slowly, while the
time-dependent phase $\Phi\left(t\right)$ varies rapidly.  So, the
Fourier transformation of the GW waveform can be calculated approximately
using the stationary phase method.  In this method, the rapidly
oscillating term is neglected and only the slowly oscillating term
survives. Then the GW waveform in the Fourier domain becomes
\begin{subequations}
\begin{align}
 &\tilde{h} \left(f\right) = \dfrac{1}{2} e^{i\Psi(t)} A(t) \left[\dfrac{2\pi}{\ddot{\Phi}(t)} \right]^{1/2}, \label{Eq:waveformf}\\
 &\Psi(t)  =  2\pi f \dfrac{D}{c} + \tilde{\Phi}(t) - \dfrac{\pi}{4}, \label{Eq:Psit_def}\\
 &\tilde{\Phi}(t) \equiv  2\pi ft - \Phi(t) , \label{Eq:tilde_Phi_def}
\end{align}
\end{subequations}
where  the time $t$ is related to frequency by $2\pi f
= \omega_{\text{gw}} \left(t\right)$.

As we will discuss in Appendix \ref{App:rewriting_waveform}, the GW
waveform Eqs. (\ref{Eq:waveformf}), (\ref{Eq:Psit_def}) and
(\ref{Eq:tilde_Phi_def}) can be rewritten explicitly in the frequency
domain as follows:
\begin{subequations}
\begin{align} 
  &\tilde{h}\left(f\right) = \mathcal{A} f^{-7/6} e^{i\Psi
 \left(f\right)} 
\chi^{19/4} 
  \left[K\left(x\right) \left( 1 + \tilde{c} J\left(x\right) \right) \right]^{-1/2}, \label{Eq:GWwaveform01}\\
  &\mathcal{A}= \left(\dfrac{5}{24} \right)^{1/2} \dfrac{1}{\pi^{2/3}}  \dfrac{c}{D} \left( \dfrac{GM_c}{c^3}\right)^{5/6}  \dfrac{1+\cos^2\iota}{2}, \label{Eq:GWwaveform02}\\
  &\Psi\left(f \right) = 2\pi f\tilde{t}_c - \Phi_c -\dfrac{\pi}{4} - \tilde{\Phi}\left(f \right), \label{Eq:GWwaveform03}\\
  &\tilde{\Phi} \left(f\right) = \dfrac{10}{3}  \left( \dfrac{8\pi GM_c}{c^3}\right)^{- 5/3} \left[
   - f \int_\infty^{f} df' \  \dfrac{\chi^{11/2}}{f'{}^{11/3} K \left( 1 + \tilde{c} J \right)  } 
   + \int_\infty^{f} df' \  \dfrac{\chi^{11/2} }{f'{}^{8/3}K \left( 1 + \tilde{c} J \right) } \right], \label{Eq:GWwaveform04}\\
  &J\left(x\right) = \dfrac{4x^{11/2-\alpha}}{\left( 1+ x^{3-\alpha}\right)^{7/2}}, \label{Eq:GWwaveform05}\\
  &K\left(x\right) = \dfrac{\left( 1 + x^{3-\alpha} \right)^{5/2} \left(1+\alpha x^{3-\alpha}/3 \right)}{1 + \left(4-\alpha\right)x^{3-\alpha}}, \label{Eq:GWwaveform06}\\
  &\chi = \left( \delta \varepsilon \right)^{1/(\alpha-3)} x, \label{Eq:GWwaveform07}\\
  &\delta = \left(\dfrac{GM_{\text{eff}}}{\pi^2 f^2} \right)^{(3-\alpha)/3}, \label{Eq:GWwaveform09}
\end{align}
\end{subequations}
where $\mathcal{A}$ is the overall amplitude, $M_c$ is the chirp mass
defined by $M_c \equiv \mu^{3/5} \meff^{5/2}$, $\tilde{t}_c$ is the
sum of the binary coalescence time $t_c$ and $D/c$, $\Phi_c$ is the
phase at coalescence, $\alpha$ is the power-law index of the DM
mini-spike, $\tilde{c}$ is defined by $\tilde{c} \equiv
c_{\text{DF}}/c_{\text{GW}}$, $\delta$ is a new frequency variable,
$x$ is defined in Eq. (\ref{Eq:def_x}), and $\varepsilon$ is defined
by Eq. (\ref{Eq:def_epsilon}).  The DM information is encoded in the
waveform Eq. (\ref{Eq:GWwaveform11}) through $K\left(x\right)$,
$J\left(x\right)$, $\chi$, $\tilde{c}$ and $M_{\text{eff}}$.  So if we
take $K\left(x\right) \to 1$, $\chi \to 1$, $\tilde{c} \to 0$,
$M_{\text{eff}} \to M_{\text{BH}}$, then Eq. (\ref{Eq:GWwaveform11})
becomes the waveform without the DM shown in
Eqs. (\ref{Eq:GW_without_DM_01})-(\ref{Eq:GW_without_DM_04}).
 
\subsection{$\delta \varepsilon $ expansion}
As we will discuss in the next section, we consider a five-year
observation by eLISA which corresponds to $f \gtrsim 10^{-3}
\text{Hz}$.  In this setup, $\delta \varepsilon \ll 1$ is satisfied.
For example, we get $\delta \varepsilon = 3.5 \times 10^{-6}$ for
$\alpha=7/3$, $f=0.01 \text{Hz}$, $\mu = 1\msol$ and the parameters
$\rhosp$, $\rsp$, and $\mbh$ listed in the table
\ref{tbl:DMParameters}.  So $\delta \varepsilon$ can be treated as a
small expansion parameter.  Since the measurement errors of the
physical parameters contained in GW are much more sensitive to the GW
phase rather than its amplitude, we expand the GW waveform $\tilde
h(f)$ up to the first order in $\delta\epsilon$ in the phase and up to
the zero-th order in the amplitude.  Using an expansion of $\chi$ in
$\delta\epsilon$,
\begin{align}
&\chi = 1+\dfrac{1}{3}\delta \varepsilon+ \dfrac{2-\a}{9}\delta^2 \varepsilon^2 + \cdots, \label{Eq:GWwaveform08}
\end{align}
the GW waveform given by Eqs. (\ref{Eq:GWwaveform01})-(\ref{Eq:GWwaveform09}) becomes  
\begin{subequations}
\begin{align}
  &\tilde{h}\left(f\right)  = \mathcal{A} f^{-7/6} e^{i\Psi \left(f\right)} L\left(f\right)^{-1/2}, \label{Eq:GWwaveform11}\\
  &\mathcal{A}= \left(\dfrac{5}{24} \right)^{1/2} \dfrac{1}{\pi^{2/3}}  \dfrac{c}{D} \left( \dfrac{GM_c}{c^3}\right)^{5/6}  \dfrac{1+\cos^2\iota}{2}, \label{Eq:GWwaveform12}\\
  &\tilde{\Phi} \left(f\right) = \dfrac{10}{3}  \left( \dfrac{8\pi GM_c}{c^3}\right)^{- 5/3} \left[
   - f \int_{f_{\rm ISCO}}^{f} df' \  f'{}^{-11/3}L^{-1}\left(f'\right)  
   + \int_{f_{\rm ISCO}}^{f} df' \  f'{}^{-8/3}L^{-1}\left(f'\right) \right], \label{Eq:GWwaveform13}\\
  &L\left(f\right) = 1 + 4c_\varepsilon \tilde{\delta}^{ (11-2\alpha) /
 \left[2\left(3-\alpha\right)\right]}
,\label{Eq:GWwaveform14} \\
  &\tilde{ \delta }  = \left(\dfrac{G}{\pi^2 f^2} \right)^{(3-\alpha)/3} , \label{Eq:GWwaveform15}\\
  &c_\varepsilon  = \meff^{\left(11-2\alpha\right)/6} \tilde{c} \varepsilon^{(11-2\alpha)/\left[2\left(3-\alpha\right)\right]},  \label{Eq:GWwaveform16}
\end{align}
\end{subequations}
where the overall amplitude $\mathcal{A}$ is defined by
Eq. (\ref{Eq:GWwaveform02}) and $\Psi\left(f\right)$ is defined by
Eq. (\ref{Eq:GWwaveform03}). The upper bound of the integration in Eq.
(\ref{Eq:GWwaveform13}), $f_{\rm ISCO} > f$ in the eLISA frequency
band, is the GW frequency when the stellar mass object enters the
innermost stable circular orbit. Hence, $\tilde\Phi(f)/(2\pi)$ is in
essence the GW cycles from the frequency $f$ to the coalescence.  The
post-Newtonian (PN) effects which are neglected in the above equations
must be taken into account in real data analysis.  However, the
frequency-dependence of the PN effect in the GW phase $\tilde{\Phi}
\left(f\right)$ differs from that of the DM effect which depends on
the power-law index $\alpha$.  So, the measurement accuracies of the
DM parameters as we will discuss later would not be affected seriously
by higher order terms in the PN expansion.

Until the previous sections, we have included both the dynamical
friction and the gravitational pull of the DM mini-spike.  It is
easily shown that the dynamical effect has much more impact on the
measurement accuracy of the DM parameters than the DM mini-spike does
\cite{2014arXiv1404.7140B}, and the above expression indeed includes
the dynamical friction but not the gravitational pull of the DM
mini-spike.  In fact, within the approximation in this subsection and
the following, the gravitational potential of the DM mini-spike shows
its signature only in the IMBH mass redefinition ($\mbh \rightarrow
M_{\rm eff}$ in $M_c$ in the above equations).  We however note that
even such a tiny effect as the gravitational pull of the DM mini-spike
do affect the detectability of GW thanks to the large number of the GW
cycles in the eLISA detection band \cite{Eda:2013gg}.

\begin{figure}[htbp]
\centering
\includegraphics[width=8cm]{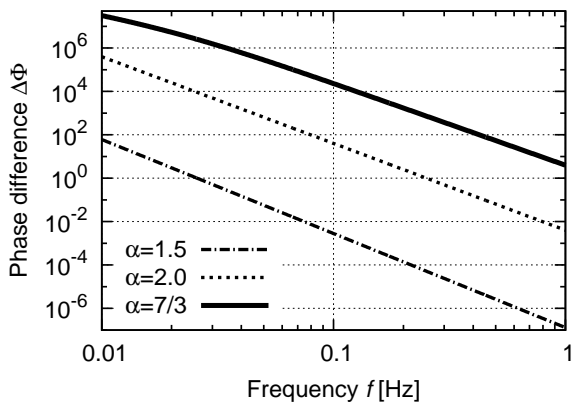}
\caption{ The accumulated phase difference $\Delta \tilde{\Phi}$
  against the power-law index $\alpha$, defined by
  Eq. (\ref{eq:phase_difference}). In essence, this is the difference
  between the accumulated phase from GW frequency $f$ and the binary
  coalescence with and without the DM mini-spike.  Three different
  curves show $\Delta \tilde \Phi$ for three different values of
  $\alpha$. For instance, if detecting a binary GW from $f=0.01$Hz to
  its coalescence, we would observe by a factor of $10^7$ more GW
  cycles in the case with a $\alpha = 7/3$ DM mini-spike than without
  any.  For this plot, we take $\mu = 1\msol$ and $\rhosp$, $\rsp$,
  and $\mbh$ are as listed in the table \ref{tbl:DMParameters}.  }
\label{Fig:phase_diff}
\end{figure}

It is important to note that the DM parameters appear only in $\alpha$
and $c_{\varepsilon}$ and that they are contained in the GW phase
$\tilde{\Phi}\left(f\right)$.  We make use of the above equations to
calculate measurement errors of the waveform parameters in the next
section.  We also define the phase difference $\Delta \tilde
\Phi\left(f\right)$ by
\begin{align}
 \Delta \tilde \Phi \left(f\right) \equiv \tilde{\Phi}\left(f\right) -
 \tilde{\Phi}_0 \left(f\right), 
\label{eq:phase_difference}
\end{align}
where $\tilde{\Phi}\left(f\right)$ defined by
Eq. (\ref{Eq:GWwaveform13}) is the phase including the DM effect and
$\tilde{\Phi}_0\left(f\right)$ defined by
Eq. (\ref{Eq:GW_without_DM_04}) is the phase without the DM
effect. $\Delta \tilde \Phi\left(f\right)$ is shown in
Fig. \ref{Fig:phase_diff} which indicates that the phase difference
becomes significant for large $\alpha$ and for the large GW frequency
$f$.  This is because in this case, the DM density near the central BH
increases and the effect of the DM on the motion of the stellar mass
object is significant.  As we discussed in our previous paper
\cite{Eda:2013gg}, the phase difference causes the mismatch between
the waveform including the DM effect and the waveform without the DM
effect.  The phase difference $\Delta \tilde{\Phi}\left(f\right)$
typically above 1 indicates the necessity to use the waveform
including the DM effect as a template.  As can be seen in
Fig. \ref{Fig:phase_diff}, if the template without the DM effect is
applied to the GW signal including the effect induced by the DM with
$\alpha > 1.5$, the resulting $S/N$ would degrade significantly.

\section{Parameter resolution for eLISA}\label{Sec:parameter_resolution}
\subsection{Brief review of the Fisher analysis}
In this subsection, we give a brief review of parameter estimation
(see \cite{Finn:1992wt,Cutler:1994ys} for more details).  Let us
consider detecting GWs with a single detector.  The detector output
$s\left(t\right)$ can be written by the sum of the GW signal
$h\left(t\right)$ and detector noise $n\left(t\right)$:
\begin{align}
 s\left(t\right) = h\left(t\right) + n\left(t\right). \label{Eq:detector_output}
\end{align} 
Assuming that the detector noise is stationary, 
the correlation between different Fourier components of the noise 
is expressed as 
\begin{align}
 \langle \tilde{n}\left(f\right) \tilde{n}^{\ast}\left(f'\right) \rangle
 = \dfrac{1}{2}\delta \left(f-f'\right)S_n\left(f\right), 
\end{align}
where the angled brackets $\langle \ \rangle$ denote an ensemble
average, the asterisk is complex conjugation and $S_n\left(f\right)$
is a one-sided power spectral density of the detector noise.  In this
paper, we consider the GW observation using eLISA which has the best
sensitivity at around $f = 0.01 \ \left[\text{Hz}\right]$.  The noise
spectral density of eLISA is given by
\begin{subequations} 
\begin{align}
  &S_n \left(f\right) = \dfrac{20}{3} \dfrac{4S_{\text{acc}} \left(f\right)/\left(2\pi f\right)^4 + S_{\text{sn}} \left(f\right) + S_{\text{omn}} \left(f\right) }{\ell^2} \left[ 1 + \left( \dfrac{f}{0.41 c/2\ell }\right) \right]^2, \label{Eq:elisa_Sn}
\end{align}
\end{subequations}
where $S_{\text{acc}} \left(f\right) = 2.13 \times 10^{-29} \left( 1 +
  10^{-4}/f \right) \left[\text{m}^2/\text{s}^{4} \text{Hz}\right]$ is
the acceleration noise spectral density, $S_{\text{sn}} \left(f\right)
= 6.28 \times 10^{-23} \left[\text{m}^2/\text{Hz}\right]$ is the shot
noise spectral density, $S_{\text{omn}} \left(f\right) =5.25 \times
10^{-23} \left[\text{m}^2/\text{Hz}\right]$ is the other measurement
noise spectral density and $\ell = 10^{9} \left[\text{m}\right]$ is
the separation between the spacecraft which is the length of its arms
of the laser interferometer (see
\cite{AmaroSeoane:2012km,AmaroSeoane:2012je} for details).

It is convenient to introduce a noise-weighted inner product between
two signals $h_1\left(t\right)$ and $h_2 \left(t\right)$ by
\begin{align}
 \left(h_1 | h_2\right) 
  &\equiv 4 \text{Re} \int_{f_{\text{ini}}}^{f_{\rm ISCO}}
   \dfrac{ \tilde{h}_1\left(f\right)\tilde{h}_2^{\ast} \left(f\right)}{ S_n\left(f\right)} \ df, \label{Eq:inner_product_def}
\end{align}
where $\text{Re} $ denotes the real part and $f_{\text{ini}}$ is the
initial frequency.  Assuming that the detector noise is Gaussian and
stationary, the probability density of the detector noise is described
by $p\left( n \right) \propto e^{-\left(n | n\right)/2}$.  We can
rewrite this expression in the form of detector signal
$s\left(t\right)$ and GWs signal $h\left(t\right)$ using
Eq. (\ref{Eq:detector_output}) as $p\left(n\right) \propto
e^{-\left(s-h|s-h \right)/2}$.

In the above case, $h\left(t\right)$ is known, while in actual GW
experiments, $h\left(t\right)$ should be replaced with a template
$h\left(t;\theta\right)$, where $\theta = \left\{ \theta_1, \cdots,
  \theta_N \right\}$ is a collection of unknown parameters.  To
determine the waveform parameters $\theta$, it is necessary to search
for the parameters which minimize the logarithm of the maximum
likelihood ratio, $\left(s-h|s-h \right) - (s|s)$.  As a result of
this process, we can infer the values of $\theta$.  However, the
expected values have  statistical errors because the detector noise
is a random process.  These measurement errors $\Delta \theta^i$ of
the waveform parameters are approximately described by the Gaussian
probability distribution for large $S/N$,
\begin{align}
 p\left( \Delta \theta^i \right) = \mathcal{N} \exp
 \left(-\dfrac{1}{2}\Gamma_{ij}\Delta\theta^i \Delta \theta^j \right),   
\end{align}
where $\mathcal{N}$ is the normalization factor and $\Gamma_{ij}$ is
called the Fisher information matrix defined by
\begin{align}
 \Gamma_{ij} \equiv \left( \dfrac{\partial h}{\partial \theta^i} \Big| \dfrac{\partial h}{\partial \theta^j} \right). \label{Eq:fisher_matrix_def}
\end{align}
The inverse of the Fisher matrix gives the root-mean-square (rms)
errors of the waveform parameters $\theta^i$:
\begin{align}
 \Delta \theta^i \equiv \sqrt{\langle \left( \Delta \theta^i\right)^2 \rangle} = \sqrt{\left(\Gamma^{-1}\right)_{ii}}, \label{Eq:rms_error}
\end{align}
where $\left(\Gamma^{-1}\right)_{ii}$ denotes the diagonal elements of
the inverse Fisher matrix.

\subsection{Preparation for parameter estimation}
The inspiral GW waveform from the IMBH surrounded by the DM mini-spike
is described by six parameters which appear in
Eqs. (\ref{Eq:GWwaveform11}) - (\ref{Eq:GWwaveform16}): the overall
amplitude, $\mathcal{A}$; the time constant, $\tilde{t}_c \equiv t_c +
D/c$, which is the sum of the traveling time $D/c$ and the coalescence
time $t_c$; the coalescence phase, $\Phi_c$; the chirp mass, $M_c$;
the two DM parameters, $\alpha$ and $c_{\varepsilon}$.  Note that the
beam pattern function of eLISA is neglected here because we are
concerned with how the DM parameters are determined by GW observations
but not with the angular resolution of eLISA (see \cite{Cutler:1997ta}
for discussion of angular resolution).

The inner product between the derivatives of the waveform with respect to
the parameters $\theta $ yields the values of the Fisher matrix
elements.  The derivatives with respect to $\mathcal{A}, \tilde{t}_c ,
\Phi_c$ and $\ln M_c$ are calculated straightforwardly as follows:
\begin{subequations}
\begin{align}
 &\dfrac{\partial \tilde{h}}{\partial \ln \mathcal{A}} = \tilde{h}, \label{Eq:waveform_derivative_01}\\
 &\dfrac{\partial \tilde{h}}{\partial \tilde{t}_c} = 2\pi i f\tilde{h}, \label{Eq:waveform_derivative_02}\\
 &\dfrac{\partial \tilde{h}}{\partial \Phi_c} = -i\tilde{h}, \label{Eq:waveform_derivative_03}\\
 &\dfrac{\partial \tilde{h}}{\partial \ln M_c} = \dfrac{5}{3}i \tilde{h} \tilde{\Phi}. \label{Eq:waveform_derivative_04}
\end{align}
\end{subequations}
The derivatives with respect to the DM parameters $\alpha,
c_{\varepsilon}$ are obtained by applying the chain rule to the
following equations:
\begin{subequations}
\begin{align}
 &\dfrac{\partial \tilde{h}}{\partial \ln \alpha}
  =\a \tilde{h}\left(
      i\dfrac{\partial \Psi}{\partial \alpha} - \dfrac{1}{2} \dfrac{1}{L}\dfrac{\partial L}{\partial \alpha}
     \right),\label{Eq:waveform_derivative_05}\\
 &\dfrac{\partial \tilde{h}}{\partial \ln c_\varepsilon}
  = c_{\varepsilon} \tilde{h}\left(
      i\dfrac{\partial \Psi}{\partial c_\varepsilon} - \dfrac{1}{2}
 \dfrac{1}{L}\dfrac{\partial L}{\partial c_\varepsilon}
     \right), \label{Eq:waveform_derivative_06}
\end{align}
\end{subequations}
where $L$ is defined in Eq. (\ref{Eq:GWwaveform14}).  However, since
the explicit expressions are complicated, we take the derivatives
numerically.

\begin{figure}[htbp]
\centering
\includegraphics[width=8cm]{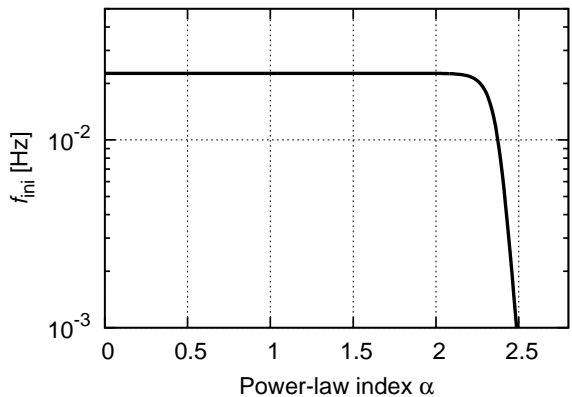}
\caption{Initial frequency against the power-law index $\alpha$.  We
  assume that the GW is detected by an eLISA 5 year observation.  For
  small $\alpha$, $f_{\text{ini}}$ is almost constant.  On the other
  hand, for large $\alpha$, $f_{\text{ini}}$ drops sharply due mainly
  to the dynamical friction.  For this plot, we take $\mu = 1\msol$
  and $\rhosp$, $\rsp$, and $\mbh$ are as listed in the table
  \ref{tbl:DMParameters}.}
\label{Fig:fini}
\end{figure}

Next, we derive the initial frequency at which the GW observation
starts.  In the presence of the DM mini-spike, the stellar mass object
orbiting the central IMBH loses its angular momentum gradually due
both to the dynamical friction and GW radiation reaction.  So the
coalescence arises earlier than the case without the DM.  Time
evolution of the frequency is described by Eq. (\ref{Eq:dfdt}),
\begin{align}
 &\dfrac{df}{d\tau} = -\dfrac{3}{5}\pi \left(\dfrac{f}{f_0} \right)^{5/3} f^2 \chi^{-11/2} \left[K \left( 1 + \tilde{c} J \right) \right],  \label{Eq:dfdtau}
\end{align}
where $f_0 \equiv c^3 / 8\pi GM_c $ and $\tau$ is the time to the
coalescence. (Note that $d\tau = -dt$.)  The lower bound
$f_{\text{ini}} \left( \alpha \right)$ of the integral in
Eq. (\ref{Eq:inner_product_def}) is required for calculating the inner
product in the Fisher matrix.  Given that the GW is observed by eLISA
for 5 years prior to the coalescence, this bound is obtained by
\begin{align}
 f_{\text{ini}} \left( \alpha \right) \equiv f\left( \alpha , \tau = 5 \ \left[ \text{yr} \right]\right).
\end{align}
By numerically solving Eq. (\ref{Eq:dfdtau}), we show the dependence
of $\alpha$ on $f_{\text{ini}}$ in Fig. \ref{Fig:fini}.  This figure
indicates that the DM mini-spike affects more strongly  the motion
of the stellar mass object for larger $\alpha$.  The initial frequency
for a 5 year observation is almost constant for small $\alpha$ due to
the smallness of the effect of the DM. Conversely, the initial
frequency drops sharply for large $\alpha$ due to the dynamical
friction from the DM.

\subsection{Measurement accuracy: The case for initially NFW profile}
In this section, we consider the most likely case that the DM
mini-spike has  an initially NFW profile, $\alpha = 7/3$.  Derivatives of
the waveform with respect to the parameters given by
Eqs. (\ref{Eq:waveform_derivative_01})-(\ref{Eq:waveform_derivative_06})
are calculated numerically.  Substitution of these results into
Eqs. (\ref{Eq:fisher_matrix_def}) and (\ref{Eq:rms_error}) gives rise
to the rms errors $\Delta \ln\mathcal{A}$, $\Delta \tilde{t}_c$,
$\Delta \Phi_c$, $\Delta \ln M_c$, $\Delta \ln\alpha$ and $\Delta \ln
c_{\varepsilon}$ as follows.
\begin{subequations}
\begin{align}
 \dfrac{\Delta \mathcal{A}}{\mathcal{A}} &= 0.1 \left(\dfrac{10}{S/N}\right), \\
 \Delta \tilde{t}_c                      &= 1.0 \left[\text{s}\right] \left(\dfrac{10}{S/N}\right) ,  \\
 \Delta \Phi_c                           &= 1.3 \left[\text{rad}\right] \left(\dfrac{10}{S/N}\right),  \\
 \dfrac{\Delta M_c}{M_c}                 &= 3.1 \times 10^{-7} \left(\dfrac{10}{S/N}\right), \label{Eq:chirpmass_error}\\
 \dfrac{\Delta \alpha}{\alpha}           &= 1.2 \times 10^{-6} \left(\dfrac{10}{S/N}\right), \label{Eq:alpha_error}\\
 \dfrac{\Delta c_\varepsilon}{c_\varepsilon} &= 5.9 \times 10^{-5} \left(\dfrac{10}{S/N}\right). 
\end{align}
\end{subequations}
Here we take $\rhosp$, $\rsp$, and $\mbh$ from  table
\ref{tbl:DMParameters} and $\mu = 1\msol$.  These measurement errors
are inversely proportional to $S/N$.  So the waveform parameters are
measurable with better accuracy for larger GW signals.  A notable
feature of the above results is that the chirp mass $M_c$ and the two
DM parameters $\alpha$ and $c_{\varepsilon}$ are determined much more
accurately than the overall amplitude $\mathcal{A}$, the coalescence
time $\tilde{t}_c$ and the coalescence phase $\Phi_c$.  This fact
reflects that $M_c$, $\alpha$ and $c_{\varepsilon}$ appear in the
phase of the waveform $\tilde{\Phi}\left(f\right)$.  From
Eqs. (\ref{Eq:number_of_circles}), (\ref{Eq:phase_t_def}) and
(\ref{Eq:tilde_Phi_def}), the GW phase is proportional to the number
of GW cycles which amplify the sensitivity to the parameters which
appear in the phase $\tilde{\Phi}\left(f\right)$ by a factor
$N_{\text{cycles}}$.  Thus, the fractional error of the chirp mass
which is proportional to the phase is order of $1/N_{\text{cycle}}$
and the two DM parameters are also determined very accurately.  In
fact, Fig. \ref{Fig:Ncycle} indicates the value of
$1/N_{\text{cycle}}$ is about $10^{-7}$, which is consistent with the
value of $\Delta M_c / M_c$ in Eq. (\ref{Eq:chirpmass_error}).

\begin{figure}[htbp]
\centering
\includegraphics[width=15cm]{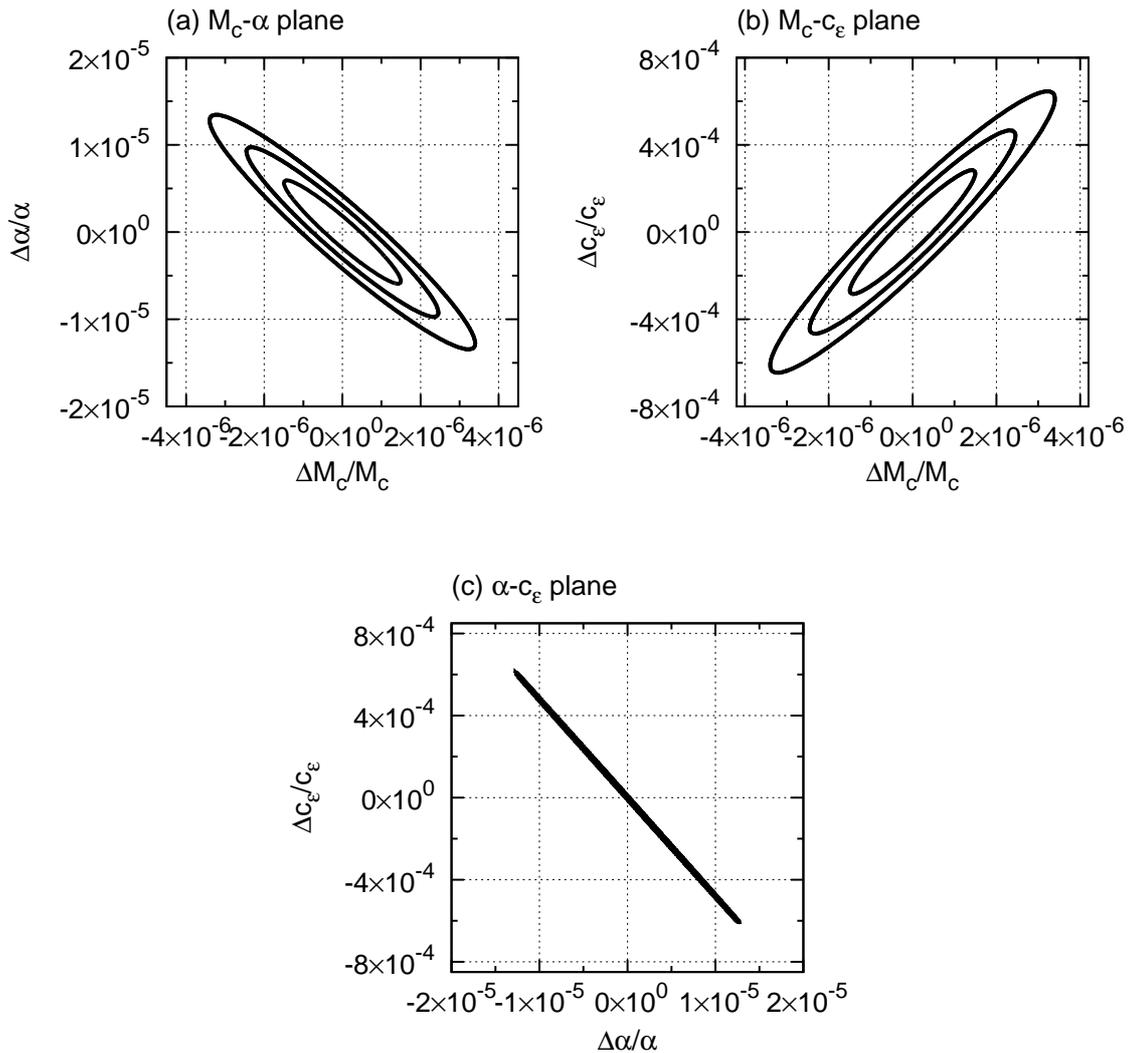}
\caption{ 
Confidence level contours of 68.3\%, 95.4\% and 99.7\% for $S/N=10$ 
in the case where the initial DM halo has
 an NFW profile and the final profile has the radial power-law index of
 $\alpha = 7/3$ through an adiabatic growth.
We assume $\rhosp$, $\rsp$, and $\mbh$ from the table \ref{tbl:DMParameters} and $\mu = 1\msol$.
}
\label{Fig:fisher_ellipse}
\end{figure}

We also investigate the correlation between the parameters which
appear in the phase, $M_c$, $\alpha $ and $c_{\varepsilon}$.
Figure. \ref{Fig:fisher_ellipse} illustrates the Fisher ellipses for
$M_c, \alpha$ and $c_{\varepsilon}$ in $S/N = 10$.  From the figures,
we observe that $M_c, \alpha$ and $c_{\varepsilon}$ are strongly
correlated with each other because all of them are contained in the
phase.  However, they are not completely degenerate and are determined
independently.  This fact can be traced to the difference of the
frequency-dependence between $M_c, \alpha$ and $c_{\varepsilon}$.

\begin{figure}[htbp]
\centering
\includegraphics[width=18cm]{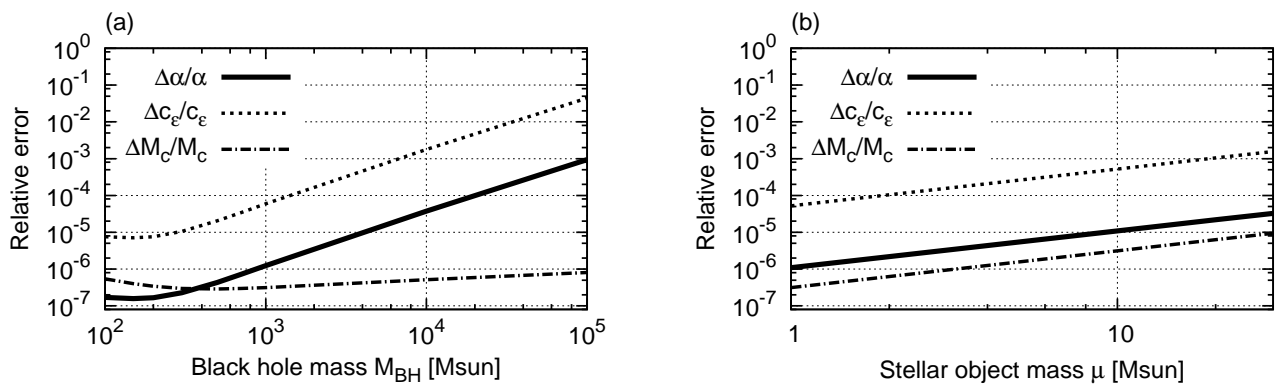}
\caption{The relative errors of the parameters in the phase
  $\tilde{\Phi}\left(f\right)$ versus (a) the central BH mass
  $M_{\text{BH}}$ and (b) the stellar mass object mass $\mu$ for $S/N
  = 10$ and $\alpha = 7/3$.  For this plot, $\rhosp$ and $\rsp$ are
  taken from the table \ref{tbl:DMParameters}.  The other parameter is
  fixed to be $\mu=1\msol$ in the left and $\mbh=10^3\msol$ in the
  right, respectively.  Note that the both axes are in the logarithmic
  scales.  The solid line, the dashed line, the dashed-dotted line
  correspond to $\Delta \alpha/\alpha, \Delta
  c_{\varepsilon}/c_{\varepsilon}, \Delta M_c/M_c$ respectively. }
\label{Fig:error_virsus_Mbh_mu}
\end{figure}

In the above discussion, the mass of the central IMBH $M_{\text{BH}}$
and that of the stellar mass object $\mu$ are fixed.  Next we analyze
the measurement errors for various values of $\mu$ and
$M_{\text{BH}}$.  The results are shown in
Fig. \ref{Fig:error_virsus_Mbh_mu}.  The figure indicates the errors
of the parameters in the phase $\tilde{\Phi}\left(f\right)$ increase
linearly with the stellar mass object mass $\mu$.  This behavior comes
from the fact that the number of cycles $N_{\text{cycle}}$ decreases
in proportion to the stellar mass object mass $\mu$.  Similarly, the
larger is the mass of the IMBH, the smaller the number of the orbital
cycles the stellar mass object experienced in the  five years prior to the
coalescence within the eLISA band. For this reason, the measurement
errors in $M_c$, $\alpha$, and $c_{\epsilon}$ increase for a larger
IMBH mass as can be seen in Fig. \ref{Fig:error_virsus_Mbh_mu}.

\subsection{Measurement accuracy: General case for initial DM profile}
\begin{figure}[htbp]
\centering
\includegraphics[width=8cm]{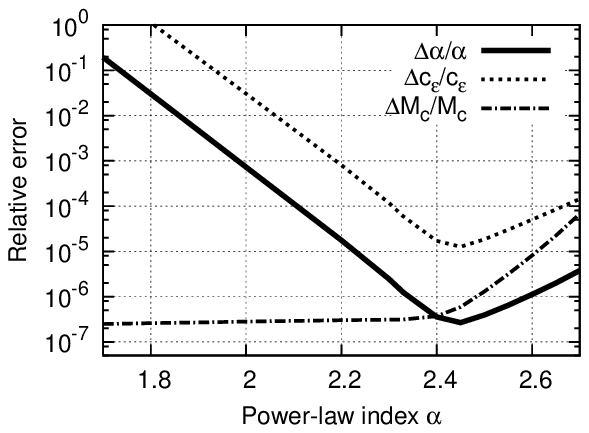}
\caption{The relative errors of the parameters in the phase
  $\tilde{\Phi}\left(f\right)$ versus the power-law index of the DM
  profile for $S/N = 10$ in the case where the DM mini-spike harboring
  the IMBH has a radially power-law profile.  The solid line, the
  dashed line, and the dashed-dotted line corresponds to $\Delta
  \alpha/\alpha, \Delta c_{\varepsilon}/c_{\varepsilon}, \Delta
  M_c/M_c$ respectively.  For this plot, $\mu = 1\msol$ and the values
  of the parameters $\mbh$, $\rhosp$, and $\rsp$ are assumed as in the
  table \ref{tbl:DMParameters}.  }
\label{Fig:error_versus_alpha}
\end{figure}

We now extend the analysis in the previous section where we considered
the case of the initially NFW profile.  We next consider the general
case without specifying the value of $\alpha$ with $\mbh$, $\rhosp$,
and $\rsp$ fixed to the values quoted in the table
\ref{tbl:DMParameters}.  The rms errors depend on the DM power-law
index $\alpha$.  We show $\Delta \ln M_c$, $\Delta \ln\alpha$ and
$\ln\Delta c_{\varepsilon}$ in Fig. \ref{Fig:error_versus_alpha}.

As shown in Fig. \ref{Fig:error_versus_alpha}, the accuracy of the DM
parameters $\Delta \ln\alpha$ and $\Delta \ln c_{\varepsilon}$ are
better for the larger $\alpha$. This is because steeper density
distributions contain more DM mass within the orbital radius (see
Fig. 2 in \cite{Eda:2013gg}).  In other words, the  steeper density
distribution has more impact on the motion of the stellar mass object
and the GW waveform is modified more strongly by the DM mini-spike.
So the DM information can be extracted from the GW waveform if the DM
mini-halo near the BH has a steep profile.
 
On the other hand, the measurement accuracy of the parameters which
appear in the phase $\tilde \Phi(f)$ become worse in $\alpha > 2.5$.
This feature can be explained by the number of GW cycles
$N_{\text{cycle}}$ which will be discussed in the Appendix
\ref{App:Ncycle}.  There we show that $N_{\text{cycle}}$ falls sharply
at $\alpha \sim 2.5$ (See Fig. \ref{Fig:Ncycle}).  The sensitivity to
the parameters which appear in the phase $\tilde \Phi(f)$ is amplified
by the number of circles $N_{\text{cycle}}$ in the frequency bandwidth
of eLISA.  For this reason, the measurement errors of $M_c, \alpha$
and $c_\varepsilon$ increase suddenly at $\alpha \sim 2.5$, as is
shown in Fig. \ref{Fig:error_versus_alpha}.  We also note that this
figure shows that we can measure the power-law index $\alpha$ at 10 \%
level even for a moderately flat radial distribution with $\alpha \sim
1.7$. In fact, when considering the gravitational pull due to the DM
potential only, it affects detectability of GW signals only for
$\alpha \gtrsim 2$. It is the dynamical friction that enables us to
explore a flatter DM distribution than ``a DM mini-spike'' referred in
the literature that has $\alpha \ge 2.25$.

\begin{figure}[htbp] 
\centering
\includegraphics[width=18cm]{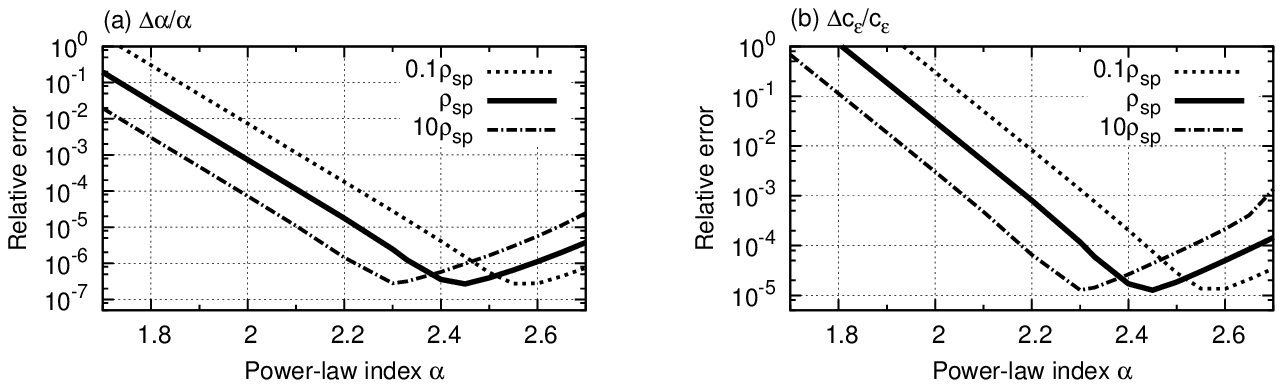}
\caption{ The relative errors of (a) $\alpha$ and (b)
  $c_{\varepsilon}$ versus the power-law index of the DM profile for
  $S/N=10$ in the case where the DM mini-spike harboring the IMBH has
  a radially power-law profile.  The solid line, the dashed line and
  the dashed-dotted line corresponds to $\rho_{\text{sp}}$, $0.1
  \times \rho_{\text{sp}}$, and $10 \times \rho_{\text{sp}}$
  respectively.  The value of $\rho_{\text{sp}}$ is taken from the
  table \ref{tbl:DMParameters}.  }
\label{Fig:error_versus_alpha_varrho}
\end{figure}

Figure \ref{Fig:error_versus_alpha_varrho} shows the relative errors
of the DM parameters, $\alpha$ and $c_{\varepsilon}$ for various
values of $\rho_{\text{sp}}$ as a function of $\alpha$.  As can be
seen in Fig. \ref{Fig:error_versus_alpha_varrho}, the relative errors
for the fixed $\alpha$ become smaller approximately linearly as the DM
density increases.  This behaviour can be traced to the amount of the
DM within the orbital radius of the stellar mass object.  It should be
noted that the value of $\rho_{\text{sp}}$ we adopt in this paper is
derived under the assumption that the initial DM mini-halo profile is
the NFW profile as discussed in the section \ref{Sec:mini-halo_model}.
Even if the DM density is an order of magnitude more sparse than that
indicated by the NFW profile, the power-law index $\alpha$ is
measurable with an accuracy of $\Delta \alpha / \alpha < 10\%$ for
$\alpha > 1.9$.
 

\section{Conclusion}\label{Sec:conclusion}
In this paper, we have investigated the measurement accuracy of dark
matter (DM) parameters by gravitational (GW) observations.  We
consider a binary system composed of an intermediate mass black hole
(IMBH) surrounded by a DM mini-spike and a stellar mass compact
object.  The compact object falling into the central IMBH is affected
by the gravitational interaction of both the IMBH and the DM
mini-spike, namely the gravitational potential of both the IMBH and the DM
mini-spike, gravitational wave back-reaction and dynamical friction.
Then the resulting inspiral GW is modified by the DM mini-spike in
comparison with the case where the IMBH has no DM mini-spike around
it.  Such a GW will be detected by future space-crafted detectors such
as eLISA/NGO.  We find that thanks to the DM parameters contained in
the GW phase, the measurement errors of the DM parameters are very
small for large power-law index of the mini-spike profile.  To put it
another way, we can extract the DM parameters very accurately from the
GW waveform using matched filtering if the DM mini-spike has a steep
profile.  Indeed, in our reference case as originally advocated
by \cite{Zhao:2005zr,Bertone:2005xz}, we could determine the power-law
index of the DM mini-spike radial profile with the 1 $\sigma$ relative
error of $\pm 5 \times 10^{-6}$ for a GW signal with
signal-to-noise-ratio 10 and assuming 5 years observation with eLISA,
as shown in Fig. \ref{Fig:fisher_ellipse} and indicated by
Eq. (\ref{Eq:alpha_error}).  We also investigated how accurately the DM
parameters can be determined for various DM parameters and the masses
of the IMBH - stellar mass object binary surrounded by a DM
mini-spike.  We have found that smaller the mass of the stellar mass
object, that of the IMBH, or the  larger the power-law index of the DM
mini-spike, we can measure DM parameters to better accuracy as shown
in Figs. \ref{Fig:error_virsus_Mbh_mu} and
\ref{Fig:error_versus_alpha}.  Even a moderately flatter mini-spike
with the radial distribution proportional to $r^{-1.7}$ would still
allow us to determine the power-law index to 10 \% accuracy.

Indirect dark matter searches in the gamma-ray band and through GW
observation proposed in our previous \cite{Eda:2013gg} and current
papers are complementary to each other. The GW observations we propose
should be applicable to both very weakly annihilating and
non-annihilating DM particles.  Even if the DM particles do not weakly
interact with each other, they affect the motion of the stellar mass
object gravitationally and the resulting GW is modified by them.  GW
are insensitive to absorption and scattering in the interstellar medium
during the propagation unlike electromagnetic waves.  Therefore GW
observations offer  information on the DM mini-spike directly.  On
the other hand, if the DM particles self-annihilate, an annihilation
plateau may develop within a Hubble time \cite{Bertone:2005hw} and the
power-law index $\alpha$ of the DM radial profile becomes effectively
zero within a radius $r_{\lim}$. For the case of the values of the
parameters listed in the table \ref{tbl:DMParameters}, the DM mass
$200$GeV, and its cross section $\sigma v = 10^{-27} {\rm cm}^3{\rm
  s}^{-1}$ \cite{Bertone:2005hw}, we find $r_{\lim} \sim
2\times10^{-4}$pc which is much larger than the initial orbital radius
at which the GW frequency from the binary enters the eLISA detection
frequency band. Hence in this case, gamma-ray searches are a better way
to explore a DM mini-halo surrounding the IMBH, as indicated by
Fig. \ref{Fig:error_versus_alpha}.

In summary, the combination of gamma-ray observations with future GW
observations will enable us to probe the structure of the DM
mini-spike and even to offer hints that may clarify the nature of DM
particles.  Moreover, because the DM profile strongly depends on the
formation history of the central IMBH, both types of observation may shed
light on how the IMBH evolved with cosmic history.

\begin{acknowledgments}
The authors thank Jun'ichi Yokoyama and Enrico Barausse for  useful comments.  
This work is supported by the Grant-in-Aid for JSPS Fellows No. 26$\cdot$8636
 (KE), the Grant-in-Aid for Young Scientists No. 25800126 (YI) 
and the MEXT Grant-in-Aid for Scientific Research on 
 Innovative Areas ``New Developments in Astrophysics Through 
 Multi-Messenger Observations of Gravitational Wave Sources'' (Grant
 Number 24103005) (YI).
\end{acknowledgments}

\appendix
\section{Rewriting the GW waveform}\label{App:rewriting_waveform}
Our goal in this appendix is to rewrite the waveform
Eq. (\ref{Eq:waveformf}) in the form of an explicit function of GW
frequency $f$.  According to Eqs. (\ref{Eq:waveformf}),
(\ref{Eq:Psit_def}) and (\ref{Eq:tilde_Phi_def}), the GW waveform from
the binary system composed of the stellar mass object and the IMBH
surrounded by DM mini-halo is expressed by
\begin{subequations}
\begin{align}
 &\tilde{h} \left(f\right) = \dfrac{1}{2} e^{i\Psi \left(t\right)} A\left(t\right) \left[\dfrac{2\pi}{\ddot{\Phi}\left(t\right)} \right]^{1/2}, \label{Eq:waveformf_Appendix}\\
 &\Psi \left(t\right) =  2\pi f
 \dfrac{\sourcedistance}{c} + \tilde{\Phi} \left(t \right) - \dfrac{\pi}{4}, \label{Eq:Psit_def_Appendix}\\
 &\tilde{\Phi} \left(t\right) \equiv 2\pi ft - \Phi \left(t\right), \label{Eq:tilde_Phi_def_Appendix}
\end{align}
\end{subequations} 
where $A\left(t\right)$ is the time-dependent amplitude defined by
Eq. (\ref{Eq:amplitude_t_def}), $\Phi\left(t\right)$ is the
time-dependent phase defined by Eq. (\ref{Eq:phase_t_def}), and
$\sourcedistance$ is the distance to the source.  We proceed as
follows.  We start with the amplitude $A/2\sqrt{2\pi / \ddot{\Phi}}$.
The frequency $f$ can be expressed in terms of the orbital radius $R$
which is related to the time $t$ by Eq. (\ref{Eq:dxdt}).  So the
amplitude can be expressed as a function of the frequency $f$ through
the relation between $t$ and $f$.  Next, we tackle the phase $\Psi$.
Finally, combining these results, we find the explicit expression of
the GW waveform in the Fourier domain.
 
\subsection{Rewriting the amplitude}
The GW frequency $f \equiv \omega_{\text{GW}} / 2\pi$ which is defined
by Eq. (\ref{Eq:orbital_frequency}) is expanded in a Taylor series in
a power of $R$:
\begin{align}
  f &= \dfrac{\omega_{\text{GW}}}{2\pi} \nonumber\\
  &= \dfrac{1}{\pi} \left[ \dfrac{GM_{\text{eff}}}{R^3} + \dfrac{F}{R^{\alpha}} \right]^{1/2} \nonumber\\
  &= \dfrac{\sqrt{G M_{\text{eff}}}}{\pi} R^{-3/2}\left[ 1 +
    \dfrac{1}{2} R^{3-\alpha} \varepsilon -\dfrac{1}{8}
    R^{2(3-\alpha)} \varepsilon^2 + \cdots \right].
\end{align}
Inverting this equation, we obtain $R$ as a function of GW frequency
and expanded in $\varepsilon$:
\begin{subequations}
\begin{align}
& R =\delta^{1/(3-\alpha)}\left[1+\dfrac{1}{3}\delta \varepsilon
        + \dfrac{2-\a}{9}\delta^2 \varepsilon^2 + \cdots \right], \\
& \delta \equiv \left(\dfrac{GM_{\text{eff}}}{\pi^2 f^2} \right)^{(3-\alpha)/3}, \label{Eq:delta_def} 
\end{align}
\end{subequations}
where we introduce a new frequency variable $\delta$ defined by
Eq. (\ref{Eq:delta_def}) for convenience.  Using the definition of $x$
given by Eq. (\ref{Eq:def_x}), the dimensionless radius parameter $x$
can be expanded in a power of $\varepsilon$:
\begin{subequations}
\begin{align} 
 &x = \left( \delta \varepsilon \right)^{1/(3-\alpha)} \chi, \label{Eq:x_delta_epsilon} \\
 &\chi \equiv 1+\dfrac{1}{3}\delta \varepsilon+ \dfrac{2-\a}{9}\delta^2 \varepsilon^2 + \cdots, \label{Eq:chi_def}
\end{align}
\end{subequations}
where we introduce $\chi$ for convenience. Note that the function
$\chi$ is equal to one when a DM mini-spike is not present around an
IMBH.
 
For later convenience, we rewrite $dx/dt$ which is defined by
Eq. $(\ref{Eq:dxdt})$ as follows.
\begin{align}
  \dfrac{dx}{dt}
  &= -c_{\text{GW}} \dfrac{\left(1+x^{3-\alpha} \right)^3}{ 4x^3\left[ 1+\left(4-\alpha \right)x^{3-\alpha} \right]} - c_{\text{DF}} \dfrac{1}{\left(1+x^{3-\alpha} \right)^{1/2}\left[ 1+\left(4-\alpha \right)x^{3-\alpha} \right]x^{-5/2 + \alpha } } \nonumber\\
  &= - c_{\text{GW}} f_{\text{GW}}\left( x\right) -  c_{\text{DF}} f_{\text{DF}}\left( x\right) \nonumber\\
  &= - c_{\text{GW}}  f_{\text{GW}}\left(x\right) \left[ 1 + \dfrac{ c_{\text{DF}}}{ c_{\text{GW}}} \dfrac{ f_{\text{DF}}\left(x\right)}{ f_{\text{GW}}\left(x\right)} \right] \nonumber\\
  &= - c_{\text{GW}}  f_{\text{GW}}\left(x\right) \left[ 1 + \tilde{c} J\left(x\right) \right], \label{Eq:dxdt_x}
\end{align}
where functions $f_{\text{GW}}\left(x\right), f_{\text{DF}}\left(x\right)$ and $J\left(x\right)$ and a coefficient $\tilde{c}$ are defined by 
\begin{subequations}
\begin{align}
  &f_{\text{GW}}\left(x\right)
    \equiv  \dfrac{\left(1+x^{3-\alpha} \right)^3}{ 4x^3\left[ 1+\left(4-\alpha \right)x^{3-\alpha} \right]},\\
  &f_{\text{DF}}\left(x\right)
    \equiv \dfrac{1}{\left(1+x^{3-\alpha} \right)^{1/2}\left[ 1+\left(4-\alpha \right)x^{3-\alpha} \right]x^{-5/2 + \alpha } },\\
  &J\left(x\right) \equiv
    \dfrac{ f_{\text{DF}}\left(x\right)}{ f_{\text{GW}}\left(x\right)} = \dfrac{4x^{11/2-\alpha}}{\left( 1+ x^{3-\alpha}\right)^{7/2}} \label{Eq:J_def},\\
  &\tilde{c} \equiv \dfrac{ c_{\text{DF}}}{ c_{\text{GW}}}.
\end{align}
\end{subequations}
The coefficient $\tilde{c}$ is the ratio of the dynamical friction
coefficient to the gravitational radiation coefficient.  So
$\tilde{c}$ includes the DM information.

Next, we rewrite the second time derivative of $\Phi$, $\ddot{\Phi}$,
as a function of $x$.  From Eq. (\ref{Eq:phase_t_def}), $\ddot{\Phi}$
is expressed by
\begin{align}
 \ddot{\Phi}\left(t\right)
 &= \dot{\omega}_{\text{GW}} \nonumber\\ 
 &= -\left(GM_{\text{eff}} \right)^{1/2}\varepsilon^{3/\left[2(3-\alpha)\right]} \dfrac{3+\alpha x^{3-\alpha}}{x^{5/2} \left(1+x^{3-\alpha}\right)^{1/2}}\dfrac{dx\left(t\right)}{dt}.\label{Eq:domegadt}
\end{align}
To move from the first line to the second, we have made use of
Eqs. (\ref{Eq:orbital_frequency}) and (\ref{Eq:def_x}).  The time
derivative of $x$ displayed in the right-hand side of
Eq. (\ref{Eq:domegadt}) can be rewritten as a function of $x$ by
Eq. (\ref{Eq:dxdt_x}).  So we can express $\ddot{\Phi}$ as a function
of $x$:
\begin{align}
 \ddot{\Phi}\left(t\right) 
 &= \left(GM_{\text{eff}} \right)^{1/2}\varepsilon^{3/\left[2(3-\alpha)\right]} c_{\text{GW}} \left[ 1 + \tilde{c} J\left(x\right) \right] \times f_{\text{GW}}\left(x\right) \dfrac{3+\alpha x^{3-\alpha}}{x^{5/2} \left(1+x^{3-\alpha}\right)^{1/2}} \nonumber\\
 &= \left(GM_{\text{eff}} \right)^{1/2}\varepsilon^{3/\left[2(3-\alpha)\right]} c_{\text{GW}} \left[ 1 + \tilde{c} J\left(x\right) \right] \times \dfrac{3}{4}x^{-11/2} 
 \dfrac{\left( 1 + x^{3-\alpha} \right)^{5/2} \left(1+\alpha x^{3-\alpha}/3 \right)}{1 + \left(4-\alpha\right)x^{3-\alpha}} \nonumber\\
 &= \left(GM_{\text{eff}} \right)^{1/2} \varepsilon^{3/\left[2(3-\alpha)\right]} c_{\text{GW}} \left[ 1 + \tilde{c} J\left(x\right) \right] \times \dfrac{3}{4}x^{-11/2} K\left(x\right), \label{Eq:ddphi_rewriting}
\end{align}
where the function $K\left(x\right)$ is defined by
\begin{align}
 K\left(x\right) \equiv \dfrac{\left( 1 + x^{3-\alpha} \right)^{5/2} \left(1+\alpha x^{3-\alpha}/3 \right)}{1 + \left(4-\alpha\right)x^{3-\alpha}}. \label{Eq:K_def}
\end{align}
Note that $K\left(x\right) $ is equal to one when a DM mini-spike is
not present around an IMBH.

Substitution of Eqs. (\ref{Eq:orbital_frequency}) and (\ref{Eq:def_x})
into E. (\ref{Eq:amplitude_t_def}) gives
\begin{align}
 A = \dfrac{4G\mu}{\sourcedistance c^4}\left(\pi f\right)^2 \varepsilon^{-2/(3-\alpha)} x^2, \label{Eq:A_rewriting}
\end{align}
after some algebra and simplification.  Combining
Eqs. (\ref{Eq:A_rewriting}) and (\ref{Eq:ddphi_rewriting}), we finally
arrive at the final expression for the amplitude,
\begin{align}
 \dfrac{A}{2}\sqrt{\dfrac{2\pi}{\ddot{\Phi}}} 
  &= \dfrac{1}{2} \times  \dfrac{4G\mu}{\sourcedistance c^4}\left(\pi f\right)^2 \varepsilon^{2/(3-\alpha)} x^2 
       \times \sqrt{\dfrac{8\pi }{3}} \left( GM_{\text{eff}}\right)^{-1/4} \varepsilon^{-3/\left[4\left( 3-\alpha\right)\right]}  c_{\text{GW}}^{-1/2}
       \left[ 1 + \tilde{c} J\left(x\right) \right]^{-1/2} x^{11/4}K\left(x\right)^{-1/2} \nonumber\\
  &= \sqrt{\dfrac{32\pi^5}{3}} \dfrac{G\mu}{\sourcedistance c^4} \left( GM_{\text{eff}}\right)^{-1/4} c_{\text{GW}}^{-1/2} \varepsilon^{5/\left[4\left( 3-\alpha\right)\right]}  
        f^{2}x^{19/4} \left[K\left(x\right) \left( 1 + \tilde{c} J\left(x\right) \right) \right]^{-1/2}  \nonumber \\
  &= \sqrt{\dfrac{5}{24}} \dfrac{1}{\pi^{2/3}}  \dfrac{c}{\sourcedistance} \left( \dfrac{GM_c}{c^3}\right)^{5/6} f^{-7/6}
     \chi^{19/4}  \left[K\left(x\right) \left( 1 + \tilde{c} J\left(x\right) \right) \right]^{-1/2}, \label{Eq:amplitude_final}
\end{align}
where $M_c$ is defined by $M_c \equiv \mu^{3/5} \meff^{5/2}$ and is
called the chirp mass.  From the second line to the third line, we
have used Eqs. (\ref{Eq:def_epsilon}), (\ref{Eq:c_GW}) and
(\ref{Eq:x_delta_epsilon}).

\subsection{Rewriting the phase}
Our next task is to express the phase $\Psi$ given by
Eq. (\ref{Eq:Psit_def_Appendix}) as a function of $x =
x\left(f\right)$.  From Eq. (\ref{Eq:ddphi_rewriting}), the time
derivative of frequency $df/dt$ is expressed by
\begin{align}
 \dfrac{df}{dt} 
  &= \dfrac{\ddot{\Phi}}{2\pi} \nonumber\\
  &= \dfrac{3}{8\pi} \left(GM_{\text{eff}} \right)^{1/2} \varepsilon^{3/\left[2(3-\alpha)\right]} c_{\text{GW}} x^{-11/2}  \left[K\left( 1 + \tilde{c} J\right) \right] \nonumber \\
  &= \dfrac{3}{5}\pi \left( \dfrac{8\pi GM_c}{c^3}\right)^{5/3} f^{11/3} \chi^{-11/2} \left[K \left( 1 + \tilde{c} J \right) \right].  \label{Eq:dfdt}
\end{align}
We used Eqs.  (\ref{Eq:c_GW}) and (\ref{Eq:x_delta_epsilon}) to go
from the second line to the third line.  Using Eq. (\ref{Eq:dfdt}), we
get
\begin{subequations}
\begin{align}
 \Phi\left(f\right) 
   &= \dfrac{10}{3}  \left( \dfrac{8\pi GM_c}{c^3}\right)^{- 5/3} \int df' \  \dfrac{\chi^{11/2}}{f'{}^{8/3}
        K \left( 1 + \tilde{c} J \right)}, \label{Eq:Phi_f}  \\
 2\pi ft
   &= - \dfrac{10}{3}  \left( \dfrac{8\pi GM_c}{c^3}\right)^{-5/3} f \int df' \  \dfrac{\chi^{11/2}}{f'{}^{11/3}
       K \left( 1 + \tilde{c} J \right)  }, \label{Eq:2pift_f}
\end{align}
\end{subequations}
where the constant of integration is determined by the initial
condition of the GW phase.

\subsection{Final form}
Collecting the above results, Eqs. (\ref{Eq:amplitude_final}),
(\ref{Eq:Phi_f}) and (\ref{Eq:2pift_f}), we finally obtain the GW
waveform in the frequency domain:
\begin{subequations}
\begin{align} 
  &\tilde{h}\left(f\right)  = \mathcal{A} f^{-7/6} e^{i\Psi \left(f\right)} \chi^{ 19/4 }  \left[K\left(x\right) \left( 1 + \tilde{c} J\left(x\right) \right) \right]^{ -1/2 },   \\
  &\mathcal{A}= \left(\dfrac{5}{24} \right)^{1/2} \dfrac{1}{\pi^{2/3}}  \dfrac{c}{\sourcedistance} \left( \dfrac{GM_c}{c^3}\right)^{5/6}  \dfrac{1+\cos^2\iota}{2}, \\
  &\Psi\left(f \right) = 2\pi f\left(t_c + \dfrac{\sourcedistance}{c}\right) - \Phi_c -\dfrac{\pi}{4} - \tilde{\Phi}\left(f \right) , \\
  &\tilde{\Phi} \left(f\right) = \dfrac{10}{3}  \left( \dfrac{8\pi GM_c}{c^3}\right)^{- 5/3} \left[
   - f \int_\infty^{f} df' \  \dfrac{\chi^{11/2}}{f'{}^{11/3} K \left( 1 + \tilde{c} J \right)  } 
   + \int_\infty^{f} df' \  \dfrac{\chi^{11/2}}{f'{}^{8/3} K \left( 1 + \tilde{c} J \right) } \right],
\end{align} 
\end{subequations}
where $\mathcal{A}$ is the overall amplitude, $t_c$ is the coalescence
time, $\Phi_c$ is the coalescence phase.  Note that when a DM
mini-spike is not present around an IMBH, $\chi \to 1$, $K \to 1$,
$M_c \to M_{c0} \equiv \mu^{3/5} \mbh^{2/5}$, so the waveform becomes
\begin{subequations}
\begin{align} 
  &\tilde{h}\left(f\right)  = \mathcal{A} f^{-7/6} e^{i\Psi \left(f\right)}, \label{Eq:GW_without_DM_01}\\
  &\mathcal{A}= \left(\dfrac{5}{24} \right)^{1/2} \dfrac{1}{\pi^{2/3}}  \dfrac{c}{\sourcedistance} \left( \dfrac{GM_{c0}}{c^3}\right)^{5/6}, \label{Eq:GW_without_DM_02}\\
  &\Psi\left(f \right) 
     = 2\pi f\left(t_c + \dfrac{\sourcedistance}{c}\right) - \Phi_c -\dfrac{\pi}{4} - \tilde{\Phi} \left(f \right), \label{Eq:GW_without_DM_03}\\
  &\tilde{\Phi} \left(f\right) =  -\dfrac{3}{4}\left( \dfrac{GM_{c0}}{c^3} 8\pi f \right)^{-5/3}. \label{Eq:GW_without_DM_04}
\end{align}
\end{subequations}
This is consistent with the waveform from the binary composed of two
point-like compact object with mass $\mu$ and $\mbh$
\cite{Cutler:1994ys}.

\section{The number of GW cycles}\label{App:Ncycle}
The detector sensitivity to the inspiral GWs is closely related to the
number of GW cycles.  That is because $N_{\text{cycle}}$ which is
defined by Eq. (\ref{Eq:number_of_circles}) is proportional to the GW
phase which is defined by Eq. (\ref{Eq:phase_t_def}).  Therefore SNR
strongly depends on the number of cycles $N_{\text{cycle}}$.  The
number of GW cycles in the frequency range $f \in \left[
  f_{\text{min}}, f_{\text{max}} \right]$ is defined by
\begin{align}
 N_{\text{cycle}}
  &= \int_{t_{\text{min}}}^{t_{\text{max}}} dt \ f\left(t\right) 
   = \int_{f_{\text{min}}}^{f_{\text{max}}} df \ \dfrac{f}{\dot{f}}, \label{Eq:number_of_circles}
\end{align}
where an overdot denotes the time derivative and $df/dt$ can be
calculated by Eq. (\ref{Eq:dfdtau}) (Note that $d\tau = -dt$.).

\begin{figure}[htbp]
\centering
\includegraphics[width=8cm]{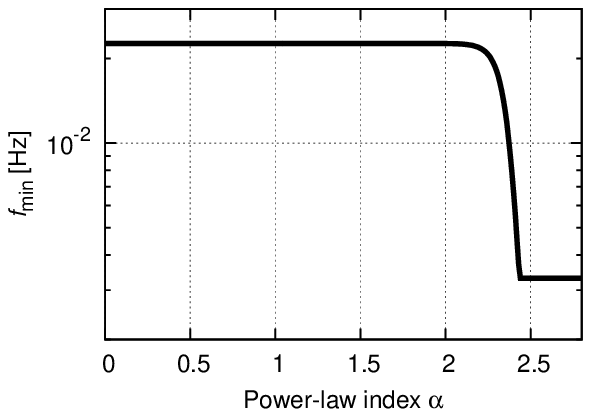}
\caption{ Lower frequency bound which is the minimum frequency the
  inspiral GW spend in the detector bandwidth $\left[f_{-}, f_{+}
  \right]$ for $\mu = 1 \msol$ and $\mbh = 10^3 \msol$.}
\label{Fig:f_min}
\end{figure}

We define the frequency bandwidth of eLISA as the frequency range $f
\in \left[f_{-}, f_{+} \right]$ within which the square root of the
noise spectral density is below half its minimum value:
\begin{align}
 \sqrt{S_n \left(f \right)} \leq 2 \sqrt{S_n \left( f_{\text{best}}
 \right)}, ~~ \left(f \in \left[f_{-}, f_{+} \right] \right),
\end{align}
where $f_{\text{best}}$ is the frequency at which the eLISA is most
sensitive to the GWs.  Because we assume 5 year observation, depending
on the binary configuration, $f_{-}$ may be smaller or larger than the
initial frequency $f_{\rm ini}$ from which the inspiral GW frequency
sweeps to the frequency at the innermost stable circular orbit,
$f_{\rm ISCO}$.  Taking the initial frequency $f_{\text{ini}}$ into
account, the frequency bandwidth $\left[f_{\text{min}}, f_{\text{max}}
\right]$ in which the inspiral GW sweeps is expressed by
\begin{subequations}
\begin{align}
 &f_{\text{min}} = \text{max} \left\{ f_{\text{ini}}, f_{-} \right\}, \label{Eq:fmin}\\
 &f_{\text{max}} = \text{min} \left\{ f_{\text{ISCO}}, f_{+} \right\}, \label{Eq:fmax}
\end{align}
\end{subequations}
and we obtain Fig. \ref{Fig:f_min} from
Eq. (\ref{Eq:number_of_circles}).  For $\alpha < 2.5$, the initial
frequency $f_{\text{ini}}$ at which the GW start to be observed is
within the full width at half minimum of $\sqrt{S_n \left(f\right)}$.
So the minimum frequency $f_{\text{min}}$ which the inspiral GW spend
in the detector bandwidth is equal to the initial frequency
$f_{\text{ini}}$.  On the other hand, for $\alpha > 2.5$, the initial
frequency $f_{\text{ini}}$ is out of the full width at half minimum.
So $f_{\text{min}}$ is equal to the lower bound of the detector
bandwidth of eLISA $f_{-}$.  Since $f_{+}$ is smaller than
$f_{\text{ISCO}}$ for all values of $\alpha$ and in the cases we
studied, $f_{\text{max}} = f_+$

\begin{figure}[htbp]
\centering
\includegraphics[width=8cm]{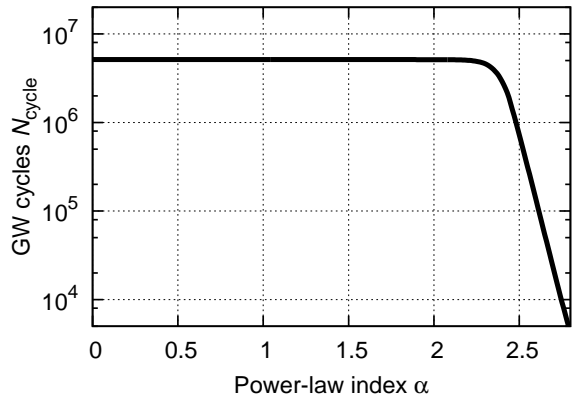}
\caption{ The number of cycles $N_{\text{cycle}}$ spent in the
  bandwidth $f \in \left[f_{\text{min}},f_{\text{max}} \right]$ for
  $\mu = 1 \msol$ and $\mbh = 10^3 \msol$.  For large $\alpha$, the
  number of cycles drops sharply because of the DM effect.}
\label{Fig:Ncycle}
\end{figure}

Using Eqs. (\ref{Eq:dfdtau}), (\ref{Eq:number_of_circles}),
(\ref{Eq:fmin}) and (\ref{Eq:fmax}), the number of cycles
$N_{\text{cycle}}$ is obtained in Fig. \ref{Fig:Ncycle}.  The figure
shows that $N_{\text{cycle}}$ is almost constant for small $\alpha$
but drops sharply for large $\alpha$.  This behavior of
$N_{\text{cycle}}$ is explained by the fact that the DM has more
influence on the motion of the stellar mass object for larger
$\alpha$.  For large $\alpha$, $df/dt$ increases sharply as $t \to
t_c$ due to the DM effect and the GW frequency of the stellar mass
object goes up rapidly through the frequency bandwidth of eLISA.  It
follows from this that larger $\alpha$ leads to wider frequency band
but to the less number of GW cycles near the best sensitivity of the
detector.  The sensitivity to the GWs is determined by the competition
between these two effects.

\end{document}